\documentclass[prl, twocolumn,amsmath,amssymb, superscriptaddress,longbibliography, nofootinbib]{revtex4-1}
\usepackage[pdftex,colorlinks=true]{hyperref}
\hypersetup{
citecolor = black
}
\usepackage{graphicx}
\usepackage{natbib,hyperref}
\usepackage{amsmath,epsfig,color,amssymb}
\usepackage{bbold}
\usepackage[dvipsnames]{xcolor}
\usepackage[export]{adjustbox}
\usepackage{comment}
\usepackage{cancel}
\newcommand{\beg}{\begin{equation}}
\newcommand{\en}{\end{equation}}

\newcommand \bel  {\begin{align}}
\newcommand \enl  {\end{align}}

\newcommand{\dg}{^\dagger}
\newcommand{\ket}[1]{|#1\rangle}

\definecolor{new}{rgb}{.08,.05,.8}

\begin{document}
\author{Ammar Kirmani}
\affiliation{Department of Physics and Astronomy, Western Washington University, Bellingham, Washington 98225, USA}
\affiliation{Physics Department, City College of the City University of New York, New York, NY 10031, USA}

\author{Kieran Bull}
\affiliation{School of Physics and Astronomy, University of Leeds, Leeds LS2 9JT, United Kingdom}

\author{Chang-Yu Hou}
\affiliation{Schlumberger-Doll Research, Cambridge, MA 02139, USA}

\author{Vedika Saravanan}
\affiliation{Department of Electrical Engineering, City College of the City University of New York, NY 10031, USA}

\author{Samah Mohamed Saeed}
\affiliation{Department of Electrical Engineering, City College of the City University of New York, NY 10031, USA}

\author{Zlatko Papi\'c}
\affiliation{School of Physics and Astronomy, University of Leeds, Leeds LS2 9JT, United Kingdom}

\author{Armin Rahmani}
\affiliation{Department of Physics and Astronomy and Advanced Materials Science and Engineering Center, Western Washington University, Bellingham, Washington 98225, USA}
\affiliation{Kavli Institute for Theoretical Physics, University of California, Santa Barbara, California 93106, USA}

\author{Pouyan Ghaemi}
\affiliation{Physics Department, City College of the City University of New York, New York, NY 10031, USA}
\affiliation{Graduate Center of the City University of New York, New York, NY 10016, USA}

\title{Probing Geometric Excitations of Fractional Quantum Hall States on Quantum Computers}

\begin{abstract}

Intermediate-scale quantum technologies provide new opportunities for scientific discovery, yet they also pose the challenge of identifying suitable problems that can take advantage of such devices in spite of their present-day limitations. 
In solid-state materials, fractional quantum Hall (FQH) phases continue to attract attention as hosts of emergent geometrical excitations analogous to gravitons, resulting from the non-perturbative interactions between the electrons. However, the direct observation of such excitations remains a challenge. Here, we identify a quasi-one-dimensional model that captures the geometric properties and graviton dynamics of FQH states. 
We then simulate geometric quench and the subsequent graviton dynamics on the IBM quantum computer using an optimally-compiled Trotter circuit with bespoke error mitigation. Moreover, we develop an efficient, optimal-control-based variational quantum algorithm that can efficiently simulate graviton dynamics in larger systems. Our results open a new avenue for studying the emergence of gravitons in a new class of tractable models on the existing quantum hardware.

\end{abstract}

\maketitle

{\bf \em Introduction.} 
While a universal fault-tolerant quantum computer with thousands of qubits remains elusive, noisy intermediate-scale quantum (NISQ) devices with a few qubits are already operational \cite{Arute2019,Boixo2018,Neill2021}, albeit with limitations due to a lack of reliable error-correction~\cite{Wootton_2020}. This progress has stirred a flurry of research activity to identify problems that can take advantage of this recently developed quantum technology
~\cite{Bharti2021}. Utilizing NISQ systems as digitized synthetic platforms to study physics phenomena challenging to investigate otherwise has emerged as a critical frontier~\cite{PhysRevApplied.9.044036}. 

In strongly-correlated electron materials, fractional quantum Hall (FQH) states are widely studied for their exotic topological properties, such as excitations with fractional charge~\cite{Laughlin1983, de-Picciotto1997} and fractional statistics~\cite{Nakamura2020, Bartolomei173}. 
Recently, FQH states have come into focus due to their universal geometric features such as Hall viscosity ~\cite{avron1995viscosity, read2009non, HaldaneViscosity} and the Girvin-MacDonald-Platzman magnetoroton collective mode~\cite{GMP85,GMP86}. In the long-wavelength limit $k\to 0$, the magnetoroton forms a quadrupole degree of freedom that carries angular momentum $L=2$ and can be represented by a quantum metric, $\widetilde g$~\cite{HaldaneGeometry}. For this reason, the $k\to 0$ limit of the magnetoroton has been referred to as ``FQH graviton"~\cite{yang2012model,Golkar2016}, due to its formal similarity with the fluctuating space-time metric in a theory of quantum gravity~\cite{bergshoeff2013zwei,bergshoeff2018gravity}. 

{The experimental detection of the FQH graviton for $\nu=1/3$ Laughlin state remains an outstanding challenge. While at large momenta, $k{\sim}\ell_B^{-1}$, with $\ell_B=\sqrt{\hbar/eB}$ being the magnetic length, the magnetoroton mode may be probed via inelastic light scattering~\cite{Pinczuk93, Platzman96, Kang01, Kukushkin09}, the magnetroroton enters the continuum  near $k\to 0$ for the $\nu=1/3$ Laughlin state (in contrast to the mode for $\nu=7/3$~\cite{refereeA1,refereeA2}).} Haldane proposed that quantum-metric fluctuations can be exposed by breaking rotational symmetry~\cite{HaldaneGeometry}. Following up on this idea, recent theoretical works~\cite{PapicMain, Lapa19} have probed the FQH graviton by quenching the metric of ``space", i.e., by suddenly making the FQH state anisotropic (see also alternative proposals~\cite{Liou19, Nguyen2021,yuzhu}). It was found that such geometric quenches induce coherent dynamics of the FQH graviton~\cite{PapicMain}, even though the graviton mode resides at finite energy densities above the FQH ground state. {\color{black} In contrast, near the FQH  liquid-nematic phase transition~\cite{Xia2011, Samkharadze2016}, the graviton is expected to emerge as a gapless excitation~\cite{PRBNem, PRXNem,PRRNem}.}

 \begin{figure}
    \centering
    \includegraphics[width=0.95\linewidth]{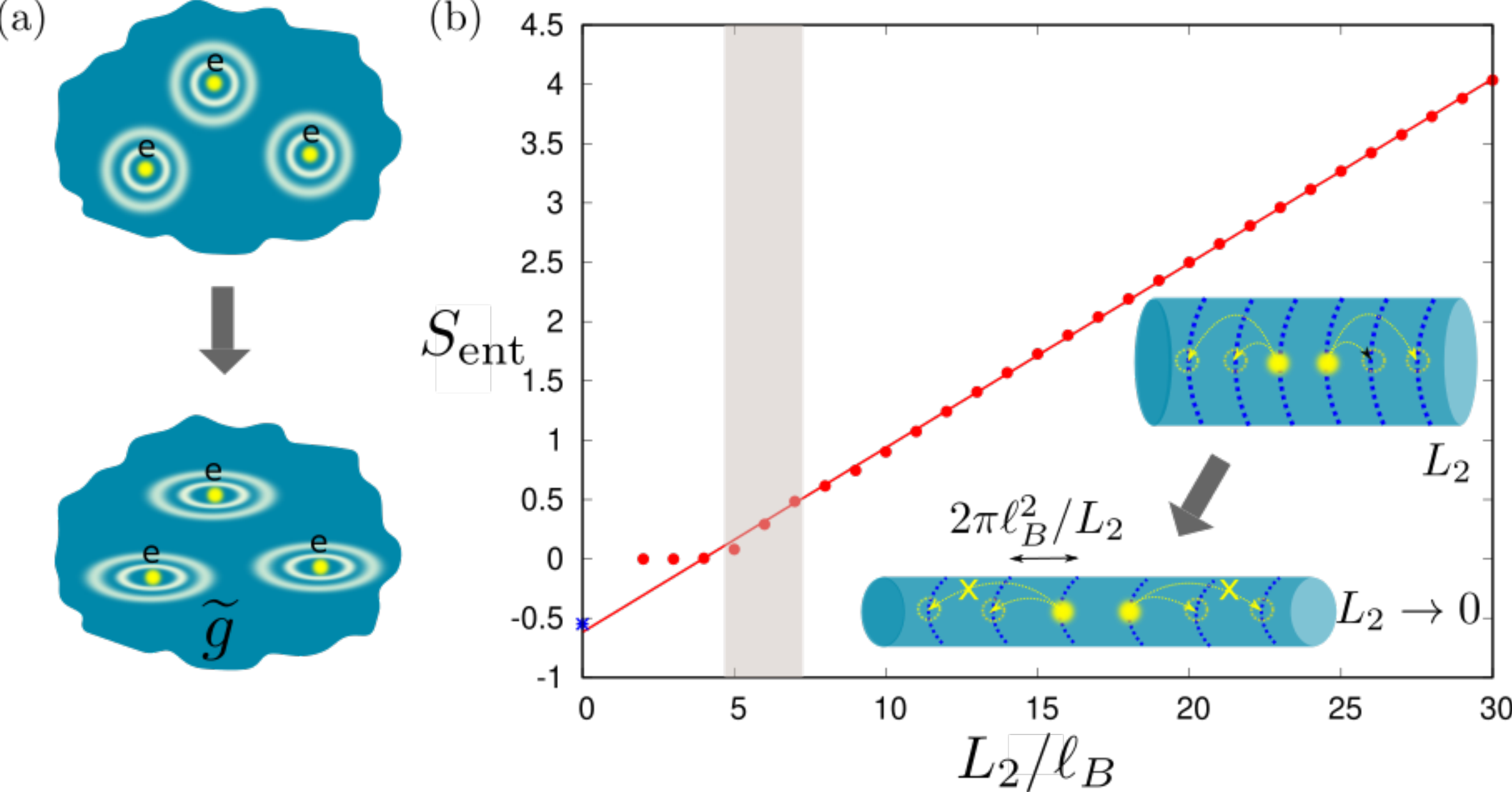}
    \caption{ (a) Geometric quench probes the fluctuations of the quantum metric $\widetilde g$~\cite{HaldaneGeometry}. 
    (b) 
    Entanglement entropy $S_\mathrm{ent}$ for the $\nu{=}1/3$ Laughlin state on a cylinder as a function of the circumference $L_2$. Entropy obeys the area law for sufficiently large circumferences $L_2 \gtrsim 5\ell_B$, with a subleading correction close to the expected value $-\ln(3)/2$ (blue star). Data is obtained using the matrix product method~\cite{Zaletel2012,Estienne2013} with entanglement truncation $P_\mathrm{max}{=}18$. Near the thin-cylinder limit (shaded), long-range electron hopping becomes strongly suppressed, as shown in the inset. 
    } 
    \label{fig:sketch}
\end{figure}

In this paper, we realize the FQH graviton in a synthetic NISQ system -- the IBM open-access digitized quantum processor -- and simulate its out-of-equilibrium dynamics.
We first map the problem onto a one-dimensional quantum spin chain, corresponding to the FQH state on a thin cylinder. While topological properties of FQH states have been extensively studied in this regime~\cite{HaldaneCyl, Bergholtz2005,Seidel2005, Bergholtz2008, Rahmani,Nakamura2012,Wang2013,Jolicoeur2012}, we show that this limit remarkably captures some \emph{geometric} properties of FQH systems, in particular their quench dynamics. As a second step, we implement the quench dynamics on the IBM NISQ device, using two complementary approaches. On the one hand, we used an optimally-compiled, noise-aware Trotterization circuit with error mitigation methods~\cite{saravanan2022pauli,Qiskit,9259942}. This allowed us to successfully simulate quench dynamics on the IBM device, overcoming the problem of the large circuit depth. 
On the other hand, we devised an efficient optimal-control-based~\cite{werschnik_quantum_2007,petersen_quantum_2010,Rahmani13} variational quantum algorithm \cite{Peruzzo:14,Wecker:15,wecker,McClean_2016}, analogous to the Quantum Approximate Optimization Algorithm (QAOA)~\cite{Farhi, Farhi:3,Yang17,Wang18,Zhou20,green1}, that creates the post-quench state using a hybrid classical-quantum approach~\cite{VQAnat1,VQAnat2}. We demonstrate that this method scales favorably with system size, with a linear-depth circuit depth and only two variational parameters.

{\bf \em Anisotropic Laughlin state {\color{black}near}  the thin-cylinder limit.}
We focus on the $\nu{=}1/3$ Laughlin FQH state~\cite{Laughlin1983} whose Hamiltonian {\it near} the thin-cylinder (TC) limit is given by~\cite{Nakamura2012}
\begin{eqnarray}\label{eq:Model}
\nonumber \hat{H} &=\sum_{j} 
V_{1,0}\hat{n}_j \hat{n}_{j+1}+V_{2,0}\hat{n}_j \hat{n}_{j+2}+V_{3,0}\hat{n}_j \hat{n}_{j+3}
\\&+ V_{2,1}c\dg_{j+1}c\dg_{j+2}c_{j+3}^{}c_{j}^{} + \mathrm{h.c.}
\end{eqnarray}
Here the operators $c_j$, $c_j^\dagger$ ($\hat n_j {\equiv} c_j^\dagger c_j$) destroy or create an electron in a Landau level (LL) orbital localized around $2\pi j \ell_B^2/L_2$. 
We assume the system is defined on a cylinder of size $L_1{\times} L_2$ containing $N$ electrons, such that the filling factor $\nu{=}N/N_\phi{=}1/3$, with magnetic flux $N_\phi{=}(L_1L_2)/(2\pi\ell_B^2)$. The {\color{black}near-TC} limit corresponds to $L_1 {\gg} L_2$ with the area ($N_\phi$) fixed, which allows us to neglect longer-range interaction terms beyond those in Eq.~(\ref{eq:Model}). 
{\color{black}Importantly, the Hamiltonian above describes a 2D system with strong spatial anisotropy, as opposed to a strictly 1D limit $L_2{\to}0$, thus allowing the emergence of the graviton mode.}
The interaction matrix elements are given by
\begin{eqnarray}\label{Vkm}
V_{k,m} =  (k^2-m^2)e^{-2\pi^2(k^2+m^2-2ikmg_{12})/{L_2^2 g_{11}}},
\end{eqnarray}
which we have generalized to the case of an arbitrary electron  mass tensor $g_{ab}$, $a,b=1,2$. The mass tensor must be symmetric and unimodular ($\mathrm{det} g{=}1$)~\cite{HaldaneGeometry}, hence we can generally write it as $g=\exp(\hat Q)$ where $\hat Q= Q (2\hat{d}_a \hat{d}_b - \delta_{a,b})$ is a Landau-de Gennes order parameter and $\hat{\mathbf{d}}=(\cos(\phi/2), \sin(\phi/2))$ is a unit vector~\cite{maciejko2013field}. Parameters $Q$ and $\phi$ intuitively represent the stretch and rotation of the metric, respectively. The FQH state is invariant under area-preserving deformations of $g$, illustrated in Fig.~\ref{fig:sketch}(a).

Since the Hamiltonian in Eq.~\eqref{eq:Model} is positive semi-definite, it has a unique (unnormalized) ground state with zero energy~\cite{Nakamura2012}
\begin{eqnarray}\label{eq:Sol}
\ket{\psi_0}= \prod_{j}\left(1- \sqrt{\frac{V_{3,0}}{V_{1,0}}} e^{i \frac{8 \pi^2 \ell_B^2}{L_2^2} \frac{g_{12}}{g_{11}}} \hat{S}_j \right) |\ldots 100100\ldots\rangle,\;\;\;\;
\end{eqnarray}
where $\hat{S}_j{=}c_{j+1}\dg c_{j+2}\dg c_{j+3} c_{j}$ is an operator that ``squeezes" two neighbouring electrons while preserving their center-of-mass position~\cite{Bernevig2008}. The ground state in the  limit $L_2{\to}0$ is the product state $|...100100...\rangle$. The off-diagonal squeezing operator is essential for the $1/3$ Laughlin state \cite{HaldaneCyl}.

In previous works~\cite{HaldaneCyl, Nakamura2012,Wang2013,Jolicoeur2012}, the ground state of the model in Eq.~\eqref{eq:Model} and its neutral excitations were studied on isotropic cylinders, 
$g_{11}{=}g_{22}{=}1$, $g_{12}{=}0$. In particular, it was found that the state in Eq.~(\ref{eq:Sol}) has  ${\sim}98\%$ overlap with the ground state of the full Hamiltonian in the range of circumferences $ 5\ell_B \lesssim L_2 \lesssim 7\ell_B$, where $V_{2,1}/V_{1,0}\approx 0.2-0.3$, justifying  the use of the truncated model Eq.~(\ref{eq:Model}) in this regime~\cite{Nakamura2012}. We have confirmed that the same conclusions continue to hold in the presence of mass anisotropy~\cite{SOM}. 

As a further justification of the model in Eq.~(\ref{eq:Model}), we plot the entanglement entropy $S_\mathrm{ent}$ of the Laughlin state in a large system of 100 electrons as a function of the circumference $L_2$ in Fig.~\ref{fig:sketch}(b). We see that it is possible to reduce $L_2$ to approximately $5\ell_B$, where the ``area law" for entanglement entropy~\cite{KitaevPreskill, LevinWen} still holds, but long-range electron hopping is strongly suppressed. Below we focus on this regime, where the key aspects of 2D physics are preserved, but the system can be mapped to a 1D spin chain model and thus efficiently simulated on quantum hardware. 

{\bf \em Geometric quench. } 
We now show that, in addition to the ground state, the effective model in Eq.~(\ref{eq:Model}) captures the high-energy excitations that govern the graviton dynamics in the FQH phase. We initially prepare the system in the ground state $|\psi_0\rangle$ in Eq.~(\ref{eq:Sol}) with isotropic metric ($g_{11}{=}g_{22}{=}1$, $g_{12}{=}0$).  At time $t{=}0$,  we instantaneously introduce diagonal anisotropy, $g_{11}'{=}1/g_{22}'{>}1$, and let the system evolve unitarily, under the dynamics 
generated by the post-quench anisotropic Hamiltonian. We are interested in the dynamical fluctuations of its quantum metric $\widetilde{g}$ as the system is taken out of equilibrium. 

Note, even though $g$ and $\widetilde g$ are related to one another, $\widetilde g$ is an emergent property of a many-body state and not necessarily equal to $g$. Nevertheless, we can formally parameterize  $\widetilde g$ using the parameters $\widetilde Q$ and $\widetilde \phi$, representing the stretch and rotation of the emergent metric. In order to determine the equations of motion for $\widetilde{Q}$ and $\widetilde{\phi}$,  we maximize the overlap between $|\psi(t)\rangle$ and the family of trial states in Eq.~(\ref{eq:Sol})~\cite{BoYangPhysRevB.85.165318}. When this overlap is close to unity, we can be confident that we found the optimal metric parameters $\widetilde{Q}$ and $\widetilde{\phi}$ describing the state $|\psi(t)\rangle$. 

 \begin{figure}[t]
	\centering
		\centering
		\includegraphics[width= \linewidth]{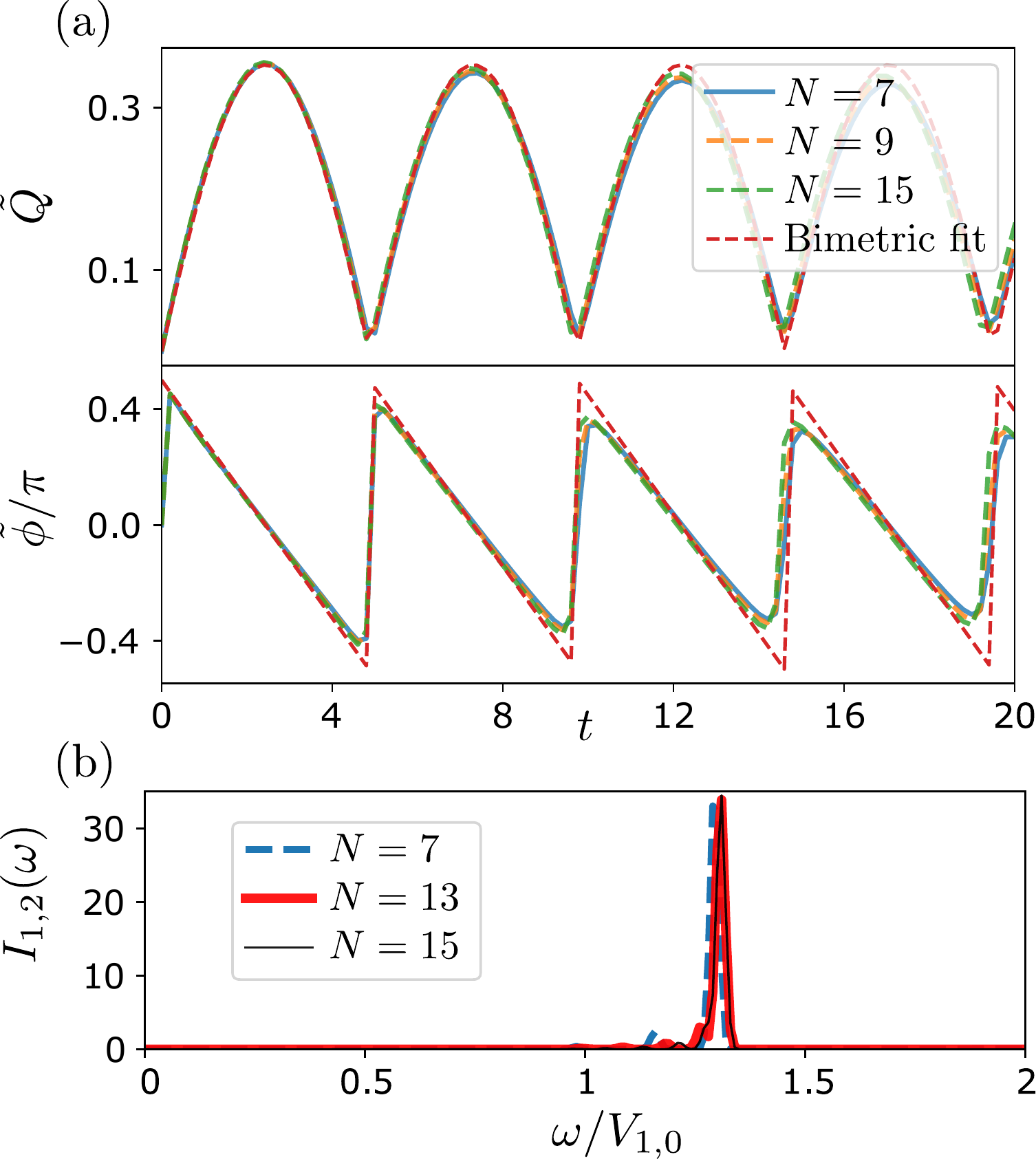}
		\caption{(a) Dynamics of $\widetilde Q$ and $\widetilde \phi$ following the geometric quench in the TC limit, with $L_2{=}5.477\ell_B$ and post-quench anisotropy $Q{\approx} 0.18$. Data is for system sizes $N{=}7, 9, 15$ electrons. (b) Quadrupole spectral function, $I_{1,2}$, shows a sharp peak at the graviton energy $E_g{\approx} 1.29$, which agrees well with the frequency of the oscillations in (a).\label{Fig:main}}
\end{figure}
In Fig.~\ref{Fig:main}, we summarize the results of the graviton dynamics in the model in Eq.~(\ref{eq:Model}) when anisotropy is suddenly changed from $Q{=}0$ to $Q{\approx} 0.18$ while keeping $\phi{=}0$. Fig.~\ref{Fig:main}(a) shows the dynamics of $\widetilde{Q}$ and $\widetilde{\phi}$ for different system sizes $N$. The dynamics is in excellent agreement with the bimetric theory in the linear regime~\cite{GromovSon}, 
\beg\label{bieqn}
 \widetilde{Q}(t)=\pm 2A \sin \frac{E_{g}t}{2},\quad \widetilde{\phi}(t)=\pm \frac{\pi}{2}-\frac{E_gt}{2},
\en
where $E_g$ is the energy of the graviton mode in units of $V_{1,0}$. As can be seen in Fig.~\ref{Fig:main}(a), the numerical data can be accurately fitted using Eqs.~(\ref{bieqn}).  
The fit yields the oscillation frequency  $E_g{=}1.29$. Note that this energy is much higher than the first excited energy of the quench Hamiltonian. We identify this energy with the graviton state as evidenced by the sharp peak in the quadrupole ($L=2$) spectral function $I_{1,2}(\omega)$~\cite{KunYang}. The later spectral function is designed to detect the characteristic $d-$wave symmetry of the graviton. 
Analogous to an oscillating space-time metric induced by a gravitational wave, $I_{1,2}(\omega)$ is the associated transition rate due to the dynamics of the oscillating mass-tensor~\cite{KunYang}. Thus, the model in Eq.~(\ref{eq:Model}) reproduces the graviton oscillation as described by the bimetric theory.

{\bf \em Spin chain mapping.}
We use the reduced registers scheme introduced in Ref.~\cite{Rahmani} to map the model \eqref{eq:Model} to a spin chain, see also~\cite{SOM} for further details. 
The reduced register is a block of three consecutive
orbitals that encodes whether or not the block is ``squeezed" with respect to the root state $|100,100,\dots \rangle$.
For each block of three sites, the state of the reduced register is $\mathbb 1$ if it is squeezed (i.e., $011$) or $\mathbb 0$ if not (i.e., either $000$ or $100$). In the root state, none of the blocks are squeezed and it maps to $|{\mathbb 0},{\mathbb 0},{\mathbb 0},\dots \rangle$. If we apply the squeezing operator to one block of the root state, we obtain, e.g., $|100,011,000,\dots \rangle\to |{\mathbb 0},{\mathbb 1},{\mathbb 0},\dots \rangle$.
In terms of reduced registers, squeezing acts as flip of $\mathbb 0$ to $ \mathbb 1$, so it can be viewed as the Pauli $X$ matrix. However, there is an important difference in that the Hilbert space is not a tensor product of reduced registers, since the squeezing can never generate two neighboring $\ldots\mathbb{1}\mathbb{1}\ldots$ configurations of the reduced registers~\cite{MoudgalyaThinTorus, MoudgalyaKrylov}. This type of constrained Hilbert space arises e.g., in the Fibonacci anyon chain~\cite{Feiguin2007}.  The inverse mapping is constructed as follows: for any $\mathbb 1$ we make a $011$ block. A $\mathbb 0$ that follows a $\mathbb 1$ ($\mathbb 0$) gives a $000$ ($100$) block. With this mapping of states, we can show that the Hamiltonian \eqref{eq:Model} maps to a local spin-chain Hamiltonian 
\begin{equation}
\label{eq:spin-Hamiltonian}
\begin{split}
    \hat{H}=&\sum_\ell \left( (V_{1,0}-3V_{3,0}){\cal N}_\ell+V_{3,0}{\cal N}_\ell{\cal N}_{\ell+2} \right.\\
    & \left. +(1-{\cal N}_{\ell-1})[{\rm Re}(V_{2,1})X_\ell-{\rm Im}(V_{2,1}) Y_\ell](1-{\cal N}_{\ell+1}) \right),
    \end{split}
\end{equation}
where we omitted the boundary terms for simplicity and introduced the occupation number $\cal N {\equiv} |\mathbb{1}\rangle \langle \mathbb{1}|$, Pauli $X {\equiv}  |\mathbb{0}\rangle \langle \mathbb{1}| {+}  |\mathbb{1}\rangle \langle \mathbb{0}|$, and Pauli $Y {\equiv}  -i |\mathbb{0}\rangle \langle \mathbb{1}| {+} i  |\mathbb{1}\rangle \langle \mathbb{0}|$ operators.

\begin{figure}[bth]
    \centering
    \includegraphics[width=1\linewidth]{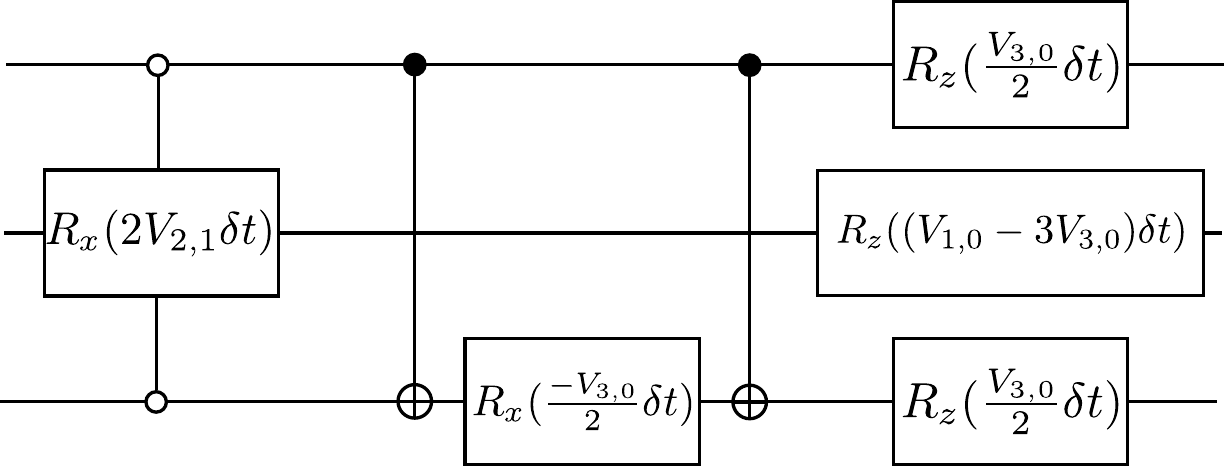}
    \caption{Circuit implementation of the trotterized unitary $U_{\ell}$ in the bulk of the spin chain.  
    }
    \label{fig:U_l}
\end{figure}

{\bf \em Quantum simulation.}
The standard procedure for simulating the time evolution $e^{-i \hat Ht}$ is to use Trotter decomposition. Here $\hat H$ is given in equation Eq.~(\ref{eq:spin-Hamiltonian}) with real $V_{2,1}$  and it has the form $\hat H=\sum_\ell H_\ell$. 
We decompose the evolution operator into $k$ Trotter steps as $e^{-i\hat Ht} \approx \left[ \prod_l U_\ell(t/k) \right]^k$, where $\delta t=t/k$ and the approximation improves for larger $k$. 
In~\cite{SOM} we derive the circuit which implements a Trotterized time evolution of our Hamiltonian and the subcircuit for the bulk $U_\ell (\delta t)$ is shown in Fig.~\ref{fig:U_l}. 
Below we demonstrate this circuit yields good results on current IBM devices with 5 qubits after using noise-aware error mitigation methods and optimized compilations~\cite{saravanan2022pauli,Qiskit,9259942}.

While the trotterization algorithm emulates the actual quantum evolution resulting from FQH quenched Hamiltonian, it has a relatively large number of entangling gates. We can access large systems by a hybrid classical-quantum method that requires classical optimization, using the following variational ansatz for the final post-quench state $|\psi(t)\rangle$:
\begin{eqnarray}\label{eq:ansatz}
 |\psi_{\rm var}(\alpha,\beta)\rangle= \prod_\ell e^{-i\alpha {\cal N}_\ell} e^{-i\beta (1-{\cal N}_{\ell-1})X_\ell}|\mathbb{000}\dots\rangle, \quad 
\end{eqnarray}
where on each reduced register $\ell$, we apply alternating gates ${\cal N}_\ell$ and $(1-{\cal N}_{\ell-1})X_\ell$ (on the very first site, due to open boundary condition, we use $X_1$ instead of $(1-{\cal N}_{\ell-1})X_\ell$ for $\ell{=}1$).

\begin{figure}[tbh]
	\centering
		\centering
		\includegraphics[width= \linewidth]{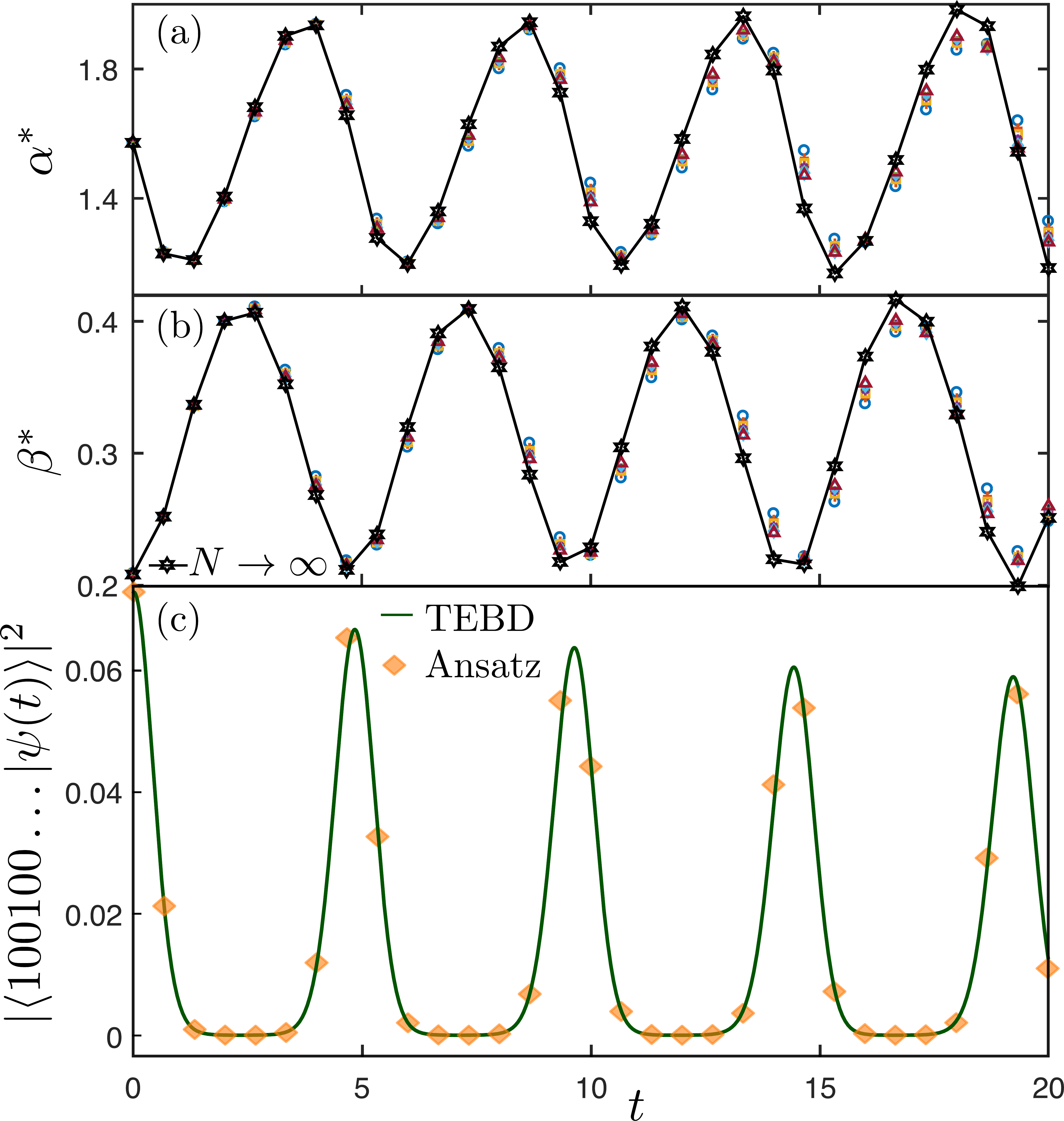}
		\caption{(a)-(b): Optimal variational parameters $\alpha^*$ and $\beta^*$ for $N=7{-}13$ particles and their extrapolation to $N{\to}\infty$ (solid black line). Optimal parameters vary smoothly with time and exhibit weak finite-size effects.
		(c) Comparison of the variational ansatz against TEBD simulation for $N{=}60$ for the overlap of the time evolved state with the root state. The parameters of the variational ansatz are extrapolated to the same system size, $N{=}60$. The ansatz with extrapolated parameters exhibits excellent agreement with the TEBD results.
		TEBD simulations were performed using a bond dimension 20 with a time step $\Delta t{=}0.01$, resulting in truncation error $10^{-5}$.
		}\label{fig:Ninf}
\end{figure}    

The optimal parameters $\alpha^*, \beta^* \in [0,2\pi)$ are determined at each time step $t$ using classical optimization by the dual annealing algorithm to maximize the overlap, $|\langle \psi_0|U^\dagger(t)|\psi_{\rm var}(\alpha,\beta)\rangle|$, with the exact state. Naively, it appears that the classical optimization needs to be performed for each $t$ and system size. Importantly, however, we find the optimal parameters $\alpha^*$, $\beta^*$ to exhibit a simple oscillatory behavior as a function of time, as well as weak dependence on the system size as shown in Fig.~\ref{fig:Ninf}(a)-(b). The data for system sizes $N=7,\dots, 13$ almost collapse, allowing a smooth extrapolation to the thermodynamic limit ($N{\to}\infty$), shown as the solid black line. In Fig.~\ref{fig:Ninf}(c), we have checked using time-evolved block decimation (TEBD)~\cite{VidalTEBD} that the extrapolated parameters produce excellent agreement with direct TEBD calculation of $|\psi(t)\rangle$ for larger systems. Thus, the weak system-size dependence of the variational parameters eliminates the need to directly perform the classical optimization for the actual size of the system, providing access to system sizes for which the classical optimization is infeasible.

{Our variational algorithm's circuit depth scales as the number of qubits $N$ independent of the evolution time $t$. As for trotterization, since we have a local lattice model in one dimension with no explicit Hamiltonian time dependence, the total circuit depth is expected to scale as $Nt$ for a fixed error tolerance \cite{Haah, Childs}. Despite higher complexity, trotterization corresponds to the actual unitary operator describing the quantum evolution and does not need any classical optimization or variational ansatz. Both algorithms have good scalability potential to more qubits. 
}

\begin{figure}[tbh]
	\centering
		\centering
		\includegraphics[width= \linewidth]{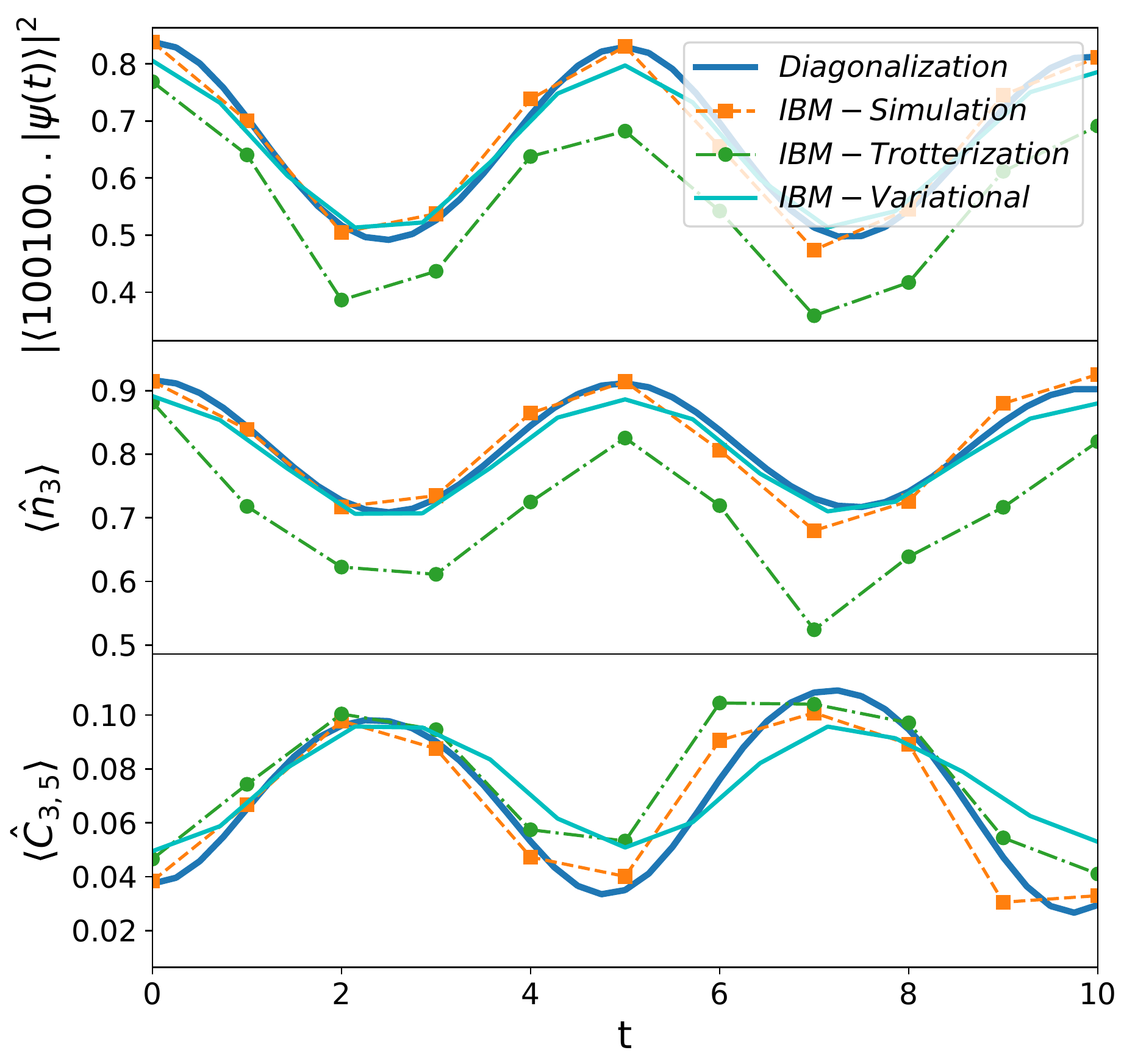}
		\caption{$N=5$ quench results for the time dependence of fidelity, density and correlation function. Comparison between exact diagonalization results, $k=15$ depth Trotterization circuit on the IBM simulator, gate optimized unitary output from IBM-Perth and variational ansatz results from IBM-Santiago. The dynamics have been simulated on IBM-Perth on date: JAN-07-2022 and on IBM-Santiago on JUN-12-2021.
		} 
		\label{fig:sammah}
\end{figure}

{\bf \em Results on the IBM Quantum Processor.}
In Fig. \ref{fig:sammah}, we present our measurements of the root state fidelity $|\langle\psi(t))|100100...\ldots\rangle|^2$, the local density $\langle n_j\rangle$ and the equal-time density-density correlation function $C_{i,j}(t)=-\langle n_i(t) n_j(t) \rangle+\langle n_i(t)\rangle \langle n_j(t) \rangle$. While these quantities are in terms of the original fermionic basis, they are easily extracted from measurements in the reduced basis using the rules discussed above Eq.~(\ref{eq:spin-Hamiltonian}). As shown in Fig.~\ref{fig:sammah}, the variational results are in excellent agreement the simulations. Similarly, the error-mitigated Trotter algorithm faithfully generates oscillations with the expected graviton frequency despite deeper circuits and higher execution-time error rates than the variational algorithm, which only induce quantitative shifts.

We note that the noise levels of the IBM devices vary widely. Using \texttt{qiskit} library, we executed error-mitigated circuit for the trottrization algorithm on \texttt{ibmq\_perth} processor~\cite{IBM} with readout error, CNOT noise and T2 dephasing time of roughly $1.4$\%, $1.7$\% and $109$ $\mu$s respectively. The  variational ansatz was executed on IBM's \texttt{ibmq\_santiago} processor~\cite{IBM} with averaged readout error, CNOT noise and T2 dephasing time of roughly $1.5$\%, $0.6$\% and $120$ $\mu$s, respectively. We also performed simulations of our circuits in \texttt{qiskit} for comparison.
Using post-selection methods, we improve the measurements by discarding states that lie outside the physical Hilbert space.

{\bf \em Conclusions.} We showed that quantum-geometrical features of FQH states can be realized in an effective 1D model that has an efficient quantum-circuit representation.  
 Our 1D model makes efficient use of resources, as each qubit corresponds to three Landau orbitals, reminiscent of holographic quantum simulation~\cite{Potter}. 
As a proof of principle, utilizing the quantum-circuit mapping, we developed efficient quantum algorithms that allowed us to simulate graviton dynamics on IBM quantum processors. {\color{black}We used state-of-the-art error mitigation to successfully run the deep trotterization circuit, which does not require any classical optimization. We also developed a variational algorithm with a linear circuit depth (independent of the evolution time), which makes use of classical optimization but can be scaled to the thermodynamic limit.} We expect these results will motivate further analytical investigations into tractable models of graviton dynamics in condensed matter systems, as well as their realizations on NISQ devices.

\section{Acknowledgements}
We thank Areg Ghazaryan and Zhao Liu for useful comments. We acknowledge the use of IBM Quantum services for this work. The views expressed are those of the authors, and do not reflect the official policy or position of IBM or the IBM Quantum team. A.K. is supported by grant numbers DMR-2037996 and DMR-2038028. P.G. acknowledges support from NSF award number  DMR-2037996 and DMR-1824265. A.R.  acknowledges support from NSF Awards DMR-1945395 and DMR-2038028. A.R. and Z.P.  thank the Kavli Institute for Theoretical Physics, acknowledging support by
the NSF under Grant PHY1748958. Z.P. and K.B. acknowledge support by EPSRC Grants No. EP/R020612/ 1 and and No. EP/M50807X/1, and by the Leverhulme Trust Research Leadership Award No. RL-2019-015. Statement of compliance with EPSRC policy framework on research data: This publication is theoretical work that does not require supporting research data. 

\bibliographystyle{apsrev4-1}
\bibliography{FQHE.bib}

\begin{thebibliography}{80}%
\makeatletter
\providecommand \@ifxundefined [1]{%
 \@ifx{#1\undefined}
}%
\providecommand \@ifnum [1]{%
 \ifnum #1\expandafter \@firstoftwo
 \else \expandafter \@secondoftwo
 \fi
}%
\providecommand \@ifx [1]{%
 \ifx #1\expandafter \@firstoftwo
 \else \expandafter \@secondoftwo
 \fi
}%
\providecommand \natexlab [1]{#1}%
\providecommand \enquote  [1]{``#1''}%
\providecommand \bibnamefont  [1]{#1}%
\providecommand \bibfnamefont [1]{#1}%
\providecommand \citenamefont [1]{#1}%
\providecommand \href@noop [0]{\@secondoftwo}%
\providecommand \href [0]{\begingroup \@sanitize@url \@href}%
\providecommand \@href[1]{\@@startlink{#1}\@@href}%
\providecommand \@@href[1]{\endgroup#1\@@endlink}%
\providecommand \@sanitize@url [0]{\catcode `\\12\catcode `\$12\catcode
  `\&12\catcode `\#12\catcode `\^12\catcode `\_12\catcode `\%12\relax}%
\providecommand \@@startlink[1]{}%
\providecommand \@@endlink[0]{}%
\providecommand \url  [0]{\begingroup\@sanitize@url \@url }%
\providecommand \@url [1]{\endgroup\@href {#1}{\urlprefix }}%
\providecommand \urlprefix  [0]{URL }%
\providecommand \Eprint [0]{\href }%
\providecommand \doibase [0]{http://dx.doi.org/}%
\providecommand \selectlanguage [0]{\@gobble}%
\providecommand \bibinfo  [0]{\@secondoftwo}%
\providecommand \bibfield  [0]{\@secondoftwo}%
\providecommand \translation [1]{[#1]}%
\providecommand \BibitemOpen [0]{}%
\providecommand \bibitemStop [0]{}%
\providecommand \bibitemNoStop [0]{.\EOS\space}%
\providecommand \EOS [0]{\spacefactor3000\relax}%
\providecommand \BibitemShut  [1]{\csname bibitem#1\endcsname}%
\let\auto@bib@innerbib\@empty
\bibitem [{\citenamefont {Arute}\ \emph {et~al.}(2019)\citenamefont {Arute}
  \emph {et~al.}}]{Arute2019}%
  \BibitemOpen
  \bibfield  {author} {\bibinfo {author} {\bibfnamefont {F.}~\bibnamefont
  {Arute}} \emph {et~al.},\ }\href {\doibase 10.1038/s41586-019-1666-5}
  {\bibfield  {journal} {\bibinfo  {journal} {Nature}\ }\textbf {\bibinfo
  {volume} {574}},\ \bibinfo {pages} {505} (\bibinfo {year}
  {2019})}\BibitemShut {NoStop}%
\bibitem [{\citenamefont {Boixo}\ \emph {et~al.}(2018)\citenamefont {Boixo},
  \citenamefont {Isakov}, \citenamefont {Smelyanskiy}, \citenamefont {Babbush},
  \citenamefont {Ding}, \citenamefont {Jiang}, \citenamefont {Bremner},
  \citenamefont {Martinis},\ and\ \citenamefont {Neven}}]{Boixo2018}%
  \BibitemOpen
  \bibfield  {author} {\bibinfo {author} {\bibfnamefont {S.}~\bibnamefont
  {Boixo}}, \bibinfo {author} {\bibfnamefont {S.~V.}\ \bibnamefont {Isakov}},
  \bibinfo {author} {\bibfnamefont {V.~N.}\ \bibnamefont {Smelyanskiy}},
  \bibinfo {author} {\bibfnamefont {R.}~\bibnamefont {Babbush}}, \bibinfo
  {author} {\bibfnamefont {N.}~\bibnamefont {Ding}}, \bibinfo {author}
  {\bibfnamefont {Z.}~\bibnamefont {Jiang}}, \bibinfo {author} {\bibfnamefont
  {M.~J.}\ \bibnamefont {Bremner}}, \bibinfo {author} {\bibfnamefont {J.~M.}\
  \bibnamefont {Martinis}}, \ and\ \bibinfo {author} {\bibfnamefont
  {H.}~\bibnamefont {Neven}},\ }\href {\doibase 10.1038/s41567-018-0124-x}
  {\bibfield  {journal} {\bibinfo  {journal} {Nature Physics}\ }\textbf
  {\bibinfo {volume} {14}},\ \bibinfo {pages} {595} (\bibinfo {year}
  {2018})}\BibitemShut {NoStop}%
\bibitem [{\citenamefont {Neill}\ \emph {et~al.}(2021)\citenamefont {Neill}
  \emph {et~al.}}]{Neill2021}%
  \BibitemOpen
  \bibfield  {author} {\bibinfo {author} {\bibfnamefont {C.}~\bibnamefont
  {Neill}} \emph {et~al.},\ }\href {\doibase 10.1038/s41586-021-03576-2}
  {\bibfield  {journal} {\bibinfo  {journal} {Nature}\ }\textbf {\bibinfo
  {volume} {594}},\ \bibinfo {pages} {508} (\bibinfo {year}
  {2021})}\BibitemShut {NoStop}%
\bibitem [{\citenamefont {Wootton}(2020)}]{Wootton_2020}%
  \BibitemOpen
  \bibfield  {author} {\bibinfo {author} {\bibfnamefont {J.~R.}\ \bibnamefont
  {Wootton}},\ }\href {\doibase 10.1088/2058-9565/aba038} {\bibfield  {journal}
  {\bibinfo  {journal} {Quantum Science and Technology}\ }\textbf {\bibinfo
  {volume} {5}},\ \bibinfo {pages} {044004} (\bibinfo {year}
  {2020})}\BibitemShut {NoStop}%
\bibitem [{\citenamefont {Bharti}\ \emph {et~al.}(2021)\citenamefont {Bharti}
  \emph {et~al.}}]{Bharti2021}%
  \BibitemOpen
  \bibfield  {author} {\bibinfo {author} {\bibfnamefont {K.}~\bibnamefont
  {Bharti}} \emph {et~al.},\ }\href@noop {} {\enquote {\bibinfo {title} {Noisy
  intermediate-scale quantum (nisq) algorithms},}\ } (\bibinfo {year} {2021}),\
  \Eprint {http://arxiv.org/abs/2101.08448} {arXiv:2101.08448 [quant-ph]}
  \BibitemShut {NoStop}%
\bibitem [{\citenamefont {Jiang}\ \emph {et~al.}(2018)\citenamefont {Jiang},
  \citenamefont {Sung}, \citenamefont {Kechedzhi}, \citenamefont
  {Smelyanskiy},\ and\ \citenamefont {Boixo}}]{PhysRevApplied.9.044036}%
  \BibitemOpen
  \bibfield  {author} {\bibinfo {author} {\bibfnamefont {Z.}~\bibnamefont
  {Jiang}}, \bibinfo {author} {\bibfnamefont {K.~J.}\ \bibnamefont {Sung}},
  \bibinfo {author} {\bibfnamefont {K.}~\bibnamefont {Kechedzhi}}, \bibinfo
  {author} {\bibfnamefont {V.~N.}\ \bibnamefont {Smelyanskiy}}, \ and\ \bibinfo
  {author} {\bibfnamefont {S.}~\bibnamefont {Boixo}},\ }\href {\doibase
  10.1103/PhysRevApplied.9.044036} {\bibfield  {journal} {\bibinfo  {journal}
  {Phys. Rev. Applied}\ }\textbf {\bibinfo {volume} {9}},\ \bibinfo {pages}
  {044036} (\bibinfo {year} {2018})}\BibitemShut {NoStop}%
\bibitem [{\citenamefont {Laughlin}(1983)}]{Laughlin1983}%
  \BibitemOpen
  \bibfield  {author} {\bibinfo {author} {\bibfnamefont {R.~B.}\ \bibnamefont
  {Laughlin}},\ }\href {\doibase 10.1103/PhysRevLett.50.1395} {\bibfield
  {journal} {\bibinfo  {journal} {Phys. Rev. Lett.}\ }\textbf {\bibinfo
  {volume} {50}},\ \bibinfo {pages} {1395} (\bibinfo {year}
  {1983})}\BibitemShut {NoStop}%
\bibitem [{\citenamefont {de~Picciotto}\ \emph {et~al.}(1997)\citenamefont
  {de~Picciotto}, \citenamefont {Reznikov}, \citenamefont {Heiblum},
  \citenamefont {Umansky}, \citenamefont {Bunin},\ and\ \citenamefont
  {Mahalu}}]{de-Picciotto1997}%
  \BibitemOpen
  \bibfield  {author} {\bibinfo {author} {\bibfnamefont {R.}~\bibnamefont
  {de~Picciotto}}, \bibinfo {author} {\bibfnamefont {M.}~\bibnamefont
  {Reznikov}}, \bibinfo {author} {\bibfnamefont {M.}~\bibnamefont {Heiblum}},
  \bibinfo {author} {\bibfnamefont {V.}~\bibnamefont {Umansky}}, \bibinfo
  {author} {\bibfnamefont {G.}~\bibnamefont {Bunin}}, \ and\ \bibinfo {author}
  {\bibfnamefont {D.}~\bibnamefont {Mahalu}},\ }\href {\doibase 10.1038/38241}
  {\bibfield  {journal} {\bibinfo  {journal} {Nature}\ }\textbf {\bibinfo
  {volume} {389}},\ \bibinfo {pages} {162} (\bibinfo {year}
  {1997})}\BibitemShut {NoStop}%
\bibitem [{\citenamefont {Nakamura}\ \emph {et~al.}(2020)\citenamefont
  {Nakamura}, \citenamefont {Liang}, \citenamefont {Gardner},\ and\
  \citenamefont {Manfra}}]{Nakamura2020}%
  \BibitemOpen
  \bibfield  {author} {\bibinfo {author} {\bibfnamefont {J.}~\bibnamefont
  {Nakamura}}, \bibinfo {author} {\bibfnamefont {S.}~\bibnamefont {Liang}},
  \bibinfo {author} {\bibfnamefont {G.~C.}\ \bibnamefont {Gardner}}, \ and\
  \bibinfo {author} {\bibfnamefont {M.~J.}\ \bibnamefont {Manfra}},\ }\href
  {\doibase 10.1038/s41567-020-1019-1} {\bibfield  {journal} {\bibinfo
  {journal} {Nature Physics}\ }\textbf {\bibinfo {volume} {16}},\ \bibinfo
  {pages} {931} (\bibinfo {year} {2020})}\BibitemShut {NoStop}%
\bibitem [{\citenamefont {Bartolomei}\ \emph {et~al.}(2020)\citenamefont
  {Bartolomei}, \citenamefont {Kumar}, \citenamefont {Bisognin}, \citenamefont
  {Marguerite}, \citenamefont {Berroir}, \citenamefont {Bocquillon},
  \citenamefont {Pla{\c c}ais}, \citenamefont {Cavanna}, \citenamefont {Dong},
  \citenamefont {Gennser}, \citenamefont {Jin},\ and\ \citenamefont
  {F{\`e}ve}}]{Bartolomei173}%
  \BibitemOpen
  \bibfield  {author} {\bibinfo {author} {\bibfnamefont {H.}~\bibnamefont
  {Bartolomei}}, \bibinfo {author} {\bibfnamefont {M.}~\bibnamefont {Kumar}},
  \bibinfo {author} {\bibfnamefont {R.}~\bibnamefont {Bisognin}}, \bibinfo
  {author} {\bibfnamefont {A.}~\bibnamefont {Marguerite}}, \bibinfo {author}
  {\bibfnamefont {J.-M.}\ \bibnamefont {Berroir}}, \bibinfo {author}
  {\bibfnamefont {E.}~\bibnamefont {Bocquillon}}, \bibinfo {author}
  {\bibfnamefont {B.}~\bibnamefont {Pla{\c c}ais}}, \bibinfo {author}
  {\bibfnamefont {A.}~\bibnamefont {Cavanna}}, \bibinfo {author} {\bibfnamefont
  {Q.}~\bibnamefont {Dong}}, \bibinfo {author} {\bibfnamefont {U.}~\bibnamefont
  {Gennser}}, \bibinfo {author} {\bibfnamefont {Y.}~\bibnamefont {Jin}}, \ and\
  \bibinfo {author} {\bibfnamefont {G.}~\bibnamefont {F{\`e}ve}},\ }\href
  {\doibase 10.1126/science.aaz5601} {\bibfield  {journal} {\bibinfo  {journal}
  {Science}\ }\textbf {\bibinfo {volume} {368}},\ \bibinfo {pages} {173}
  (\bibinfo {year} {2020})}\BibitemShut {NoStop}%
\bibitem [{\citenamefont {Avron}\ \emph {et~al.}(1995)\citenamefont {Avron},
  \citenamefont {Seiler},\ and\ \citenamefont {Zograf}}]{avron1995viscosity}%
  \BibitemOpen
  \bibfield  {author} {\bibinfo {author} {\bibfnamefont {J.~E.}\ \bibnamefont
  {Avron}}, \bibinfo {author} {\bibfnamefont {R.}~\bibnamefont {Seiler}}, \
  and\ \bibinfo {author} {\bibfnamefont {P.~G.}\ \bibnamefont {Zograf}},\
  }\href {\doibase 10.1103/PhysRevLett.75.697} {\bibfield  {journal} {\bibinfo
  {journal} {Phys. Rev. Lett.}\ }\textbf {\bibinfo {volume} {75}},\ \bibinfo
  {pages} {697} (\bibinfo {year} {1995})}\BibitemShut {NoStop}%
\bibitem [{\citenamefont {Read}(2009)}]{read2009non}%
  \BibitemOpen
  \bibfield  {author} {\bibinfo {author} {\bibfnamefont {N.}~\bibnamefont
  {Read}},\ }\href {\doibase 10.1103/PhysRevB.79.045308} {\bibfield  {journal}
  {\bibinfo  {journal} {Phys. Rev. B}\ }\textbf {\bibinfo {volume} {79}},\
  \bibinfo {pages} {045308} (\bibinfo {year} {2009})}\BibitemShut {NoStop}%
\bibitem [{\citenamefont {{Haldane}}(2009)}]{HaldaneViscosity}%
  \BibitemOpen
  \bibfield  {author} {\bibinfo {author} {\bibfnamefont {F.~D.~M.}\
  \bibnamefont {{Haldane}}},\ }\href@noop {} {\bibfield  {journal} {\bibinfo
  {journal} {ArXiv e-prints}\ } (\bibinfo {year} {2009})},\ \Eprint
  {http://arxiv.org/abs/0906.1854} {arXiv:0906.1854 [cond-mat.str-el]}
  \BibitemShut {NoStop}%
\bibitem [{\citenamefont {Girvin}\ \emph {et~al.}(1985)\citenamefont {Girvin},
  \citenamefont {MacDonald},\ and\ \citenamefont {Platzman}}]{GMP85}%
  \BibitemOpen
  \bibfield  {author} {\bibinfo {author} {\bibfnamefont {S.~M.}\ \bibnamefont
  {Girvin}}, \bibinfo {author} {\bibfnamefont {A.~H.}\ \bibnamefont
  {MacDonald}}, \ and\ \bibinfo {author} {\bibfnamefont {P.~M.}\ \bibnamefont
  {Platzman}},\ }\href {\doibase 10.1103/PhysRevLett.54.581} {\bibfield
  {journal} {\bibinfo  {journal} {Phys. Rev. Lett.}\ }\textbf {\bibinfo
  {volume} {54}},\ \bibinfo {pages} {581} (\bibinfo {year} {1985})}\BibitemShut
  {NoStop}%
\bibitem [{\citenamefont {Girvin}\ \emph {et~al.}(1986)\citenamefont {Girvin},
  \citenamefont {MacDonald},\ and\ \citenamefont {Platzman}}]{GMP86}%
  \BibitemOpen
  \bibfield  {author} {\bibinfo {author} {\bibfnamefont {S.~M.}\ \bibnamefont
  {Girvin}}, \bibinfo {author} {\bibfnamefont {A.~H.}\ \bibnamefont
  {MacDonald}}, \ and\ \bibinfo {author} {\bibfnamefont {P.~M.}\ \bibnamefont
  {Platzman}},\ }\href {\doibase 10.1103/PhysRevB.33.2481} {\bibfield
  {journal} {\bibinfo  {journal} {Phys. Rev. B}\ }\textbf {\bibinfo {volume}
  {33}},\ \bibinfo {pages} {2481} (\bibinfo {year} {1986})}\BibitemShut
  {NoStop}%
\bibitem [{\citenamefont {Haldane}(2011)}]{HaldaneGeometry}%
  \BibitemOpen
  \bibfield  {author} {\bibinfo {author} {\bibfnamefont {F.~D.~M.}\
  \bibnamefont {Haldane}},\ }\href {\doibase 10.1103/PhysRevLett.107.116801}
  {\bibfield  {journal} {\bibinfo  {journal} {Phys. Rev. Lett.}\ }\textbf
  {\bibinfo {volume} {107}},\ \bibinfo {pages} {116801} (\bibinfo {year}
  {2011})}\BibitemShut {NoStop}%
\bibitem [{\citenamefont {Yang}\ \emph
  {et~al.}(2012{\natexlab{a}})\citenamefont {Yang}, \citenamefont {Hu},
  \citenamefont {Papi\ifmmode~\acute{c}\else \'{c}\fi{}},\ and\ \citenamefont
  {Haldane}}]{yang2012model}%
  \BibitemOpen
  \bibfield  {author} {\bibinfo {author} {\bibfnamefont {B.}~\bibnamefont
  {Yang}}, \bibinfo {author} {\bibfnamefont {Z.-X.}\ \bibnamefont {Hu}},
  \bibinfo {author} {\bibfnamefont {Z.}~\bibnamefont
  {Papi\ifmmode~\acute{c}\else \'{c}\fi{}}}, \ and\ \bibinfo {author}
  {\bibfnamefont {F.~D.~M.}\ \bibnamefont {Haldane}},\ }\href {\doibase
  10.1103/PhysRevLett.108.256807} {\bibfield  {journal} {\bibinfo  {journal}
  {Phys. Rev. Lett.}\ }\textbf {\bibinfo {volume} {108}},\ \bibinfo {pages}
  {256807} (\bibinfo {year} {2012}{\natexlab{a}})}\BibitemShut {NoStop}%
\bibitem [{\citenamefont {Golkar}\ \emph {et~al.}(2016)\citenamefont {Golkar},
  \citenamefont {Nguyen},\ and\ \citenamefont {Son}}]{Golkar2016}%
  \BibitemOpen
  \bibfield  {author} {\bibinfo {author} {\bibfnamefont {S.}~\bibnamefont
  {Golkar}}, \bibinfo {author} {\bibfnamefont {D.~X.}\ \bibnamefont {Nguyen}},
  \ and\ \bibinfo {author} {\bibfnamefont {D.~T.}\ \bibnamefont {Son}},\ }\href
  {\doibase 10.1007/JHEP01(2016)021} {\bibfield  {journal} {\bibinfo  {journal}
  {Journal of High Energy Physics}\ }\textbf {\bibinfo {volume} {2016}},\
  \bibinfo {pages} {21} (\bibinfo {year} {2016})}\BibitemShut {NoStop}%
\bibitem [{\citenamefont {Bergshoeff}\ \emph {et~al.}(2013)\citenamefont
  {Bergshoeff}, \citenamefont {de~Haan}, \citenamefont {Hohm}, \citenamefont
  {Merbis},\ and\ \citenamefont {Townsend}}]{bergshoeff2013zwei}%
  \BibitemOpen
  \bibfield  {author} {\bibinfo {author} {\bibfnamefont {E.~A.}\ \bibnamefont
  {Bergshoeff}}, \bibinfo {author} {\bibfnamefont {S.}~\bibnamefont {de~Haan}},
  \bibinfo {author} {\bibfnamefont {O.}~\bibnamefont {Hohm}}, \bibinfo {author}
  {\bibfnamefont {W.}~\bibnamefont {Merbis}}, \ and\ \bibinfo {author}
  {\bibfnamefont {P.~K.}\ \bibnamefont {Townsend}},\ }\href {\doibase
  10.1103/PhysRevLett.111.111102} {\bibfield  {journal} {\bibinfo  {journal}
  {Phys. Rev. Lett.}\ }\textbf {\bibinfo {volume} {111}},\ \bibinfo {pages}
  {111102} (\bibinfo {year} {2013})}\BibitemShut {NoStop}%
\bibitem [{\citenamefont {Bergshoeff}\ \emph {et~al.}(2018)\citenamefont
  {Bergshoeff}, \citenamefont {Rosseel},\ and\ \citenamefont
  {Townsend}}]{bergshoeff2018gravity}%
  \BibitemOpen
  \bibfield  {author} {\bibinfo {author} {\bibfnamefont {E.~A.}\ \bibnamefont
  {Bergshoeff}}, \bibinfo {author} {\bibfnamefont {J.}~\bibnamefont {Rosseel}},
  \ and\ \bibinfo {author} {\bibfnamefont {P.~K.}\ \bibnamefont {Townsend}},\
  }\href@noop {} {\bibfield  {journal} {\bibinfo  {journal} {Physical review
  letters}\ }\textbf {\bibinfo {volume} {120}},\ \bibinfo {pages} {141601}
  (\bibinfo {year} {2018})}\BibitemShut {NoStop}%
\bibitem [{\citenamefont {Pinczuk}\ \emph {et~al.}(1993)\citenamefont
  {Pinczuk}, \citenamefont {Dennis}, \citenamefont {Pfeiffer},\ and\
  \citenamefont {West}}]{Pinczuk93}%
  \BibitemOpen
  \bibfield  {author} {\bibinfo {author} {\bibfnamefont {A.}~\bibnamefont
  {Pinczuk}}, \bibinfo {author} {\bibfnamefont {B.~S.}\ \bibnamefont {Dennis}},
  \bibinfo {author} {\bibfnamefont {L.~N.}\ \bibnamefont {Pfeiffer}}, \ and\
  \bibinfo {author} {\bibfnamefont {K.}~\bibnamefont {West}},\ }\href {\doibase
  10.1103/PhysRevLett.70.3983} {\bibfield  {journal} {\bibinfo  {journal}
  {Phys. Rev. Lett.}\ }\textbf {\bibinfo {volume} {70}},\ \bibinfo {pages}
  {3983} (\bibinfo {year} {1993})}\BibitemShut {NoStop}%
\bibitem [{\citenamefont {Platzman}\ and\ \citenamefont
  {He}(1996)}]{Platzman96}%
  \BibitemOpen
  \bibfield  {author} {\bibinfo {author} {\bibfnamefont {P.~M.}\ \bibnamefont
  {Platzman}}\ and\ \bibinfo {author} {\bibfnamefont {S.}~\bibnamefont {He}},\
  }\href {\doibase 10.1088/0031-8949/1996/t66/030} {\bibfield  {journal}
  {\bibinfo  {journal} {Phys. Scr.}\ }\textbf {\bibinfo {volume} {T66}},\
  \bibinfo {pages} {167} (\bibinfo {year} {1996})}\BibitemShut {NoStop}%
\bibitem [{\citenamefont {Kang}\ \emph {et~al.}(2001)\citenamefont {Kang},
  \citenamefont {Pinczuk}, \citenamefont {Dennis}, \citenamefont {Pfeiffer},\
  and\ \citenamefont {West}}]{Kang01}%
  \BibitemOpen
  \bibfield  {author} {\bibinfo {author} {\bibfnamefont {M.}~\bibnamefont
  {Kang}}, \bibinfo {author} {\bibfnamefont {A.}~\bibnamefont {Pinczuk}},
  \bibinfo {author} {\bibfnamefont {B.~S.}\ \bibnamefont {Dennis}}, \bibinfo
  {author} {\bibfnamefont {L.~N.}\ \bibnamefont {Pfeiffer}}, \ and\ \bibinfo
  {author} {\bibfnamefont {K.~W.}\ \bibnamefont {West}},\ }\href {\doibase
  10.1103/PhysRevLett.86.2637} {\bibfield  {journal} {\bibinfo  {journal}
  {Phys. Rev. Lett.}\ }\textbf {\bibinfo {volume} {86}},\ \bibinfo {pages}
  {2637} (\bibinfo {year} {2001})}\BibitemShut {NoStop}%
\bibitem [{\citenamefont {Kukushkin}\ \emph {et~al.}(2009)\citenamefont
  {Kukushkin}, \citenamefont {Smet}, \citenamefont {Scarola}, \citenamefont
  {Umansky},\ and\ \citenamefont {von Klitzing}}]{Kukushkin09}%
  \BibitemOpen
  \bibfield  {author} {\bibinfo {author} {\bibfnamefont {I.~V.}\ \bibnamefont
  {Kukushkin}}, \bibinfo {author} {\bibfnamefont {J.~H.}\ \bibnamefont {Smet}},
  \bibinfo {author} {\bibfnamefont {V.~W.}\ \bibnamefont {Scarola}}, \bibinfo
  {author} {\bibfnamefont {V.}~\bibnamefont {Umansky}}, \ and\ \bibinfo
  {author} {\bibfnamefont {K.}~\bibnamefont {von Klitzing}},\ }\href {\doibase
  10.1126/science.1171472} {\bibfield  {journal} {\bibinfo  {journal}
  {Science}\ }\textbf {\bibinfo {volume} {324}},\ \bibinfo {pages} {1044}
  (\bibinfo {year} {2009})}\BibitemShut {NoStop}%
\bibitem [{\citenamefont {Wurstbauer}\ \emph {et~al.}(2015)\citenamefont
  {Wurstbauer}, \citenamefont {Levy}, \citenamefont {Pinczuk}, \citenamefont
  {West}, \citenamefont {Pfeiffer}, \citenamefont {Manfra}, \citenamefont
  {Gardner},\ and\ \citenamefont {Watson}}]{refereeA1}%
  \BibitemOpen
  \bibfield  {author} {\bibinfo {author} {\bibfnamefont {U.}~\bibnamefont
  {Wurstbauer}}, \bibinfo {author} {\bibfnamefont {A.~L.}\ \bibnamefont
  {Levy}}, \bibinfo {author} {\bibfnamefont {A.}~\bibnamefont {Pinczuk}},
  \bibinfo {author} {\bibfnamefont {K.~W.}\ \bibnamefont {West}}, \bibinfo
  {author} {\bibfnamefont {L.~N.}\ \bibnamefont {Pfeiffer}}, \bibinfo {author}
  {\bibfnamefont {M.~J.}\ \bibnamefont {Manfra}}, \bibinfo {author}
  {\bibfnamefont {G.~C.}\ \bibnamefont {Gardner}}, \ and\ \bibinfo {author}
  {\bibfnamefont {J.~D.}\ \bibnamefont {Watson}},\ }\href {\doibase
  10.1103/PhysRevB.92.241407} {\bibfield  {journal} {\bibinfo  {journal} {Phys.
  Rev. B}\ }\textbf {\bibinfo {volume} {92}},\ \bibinfo {pages} {241407}
  (\bibinfo {year} {2015})}\BibitemShut {NoStop}%
\bibitem [{\citenamefont {Jolicoeur}(2017)}]{refereeA2}%
  \BibitemOpen
  \bibfield  {author} {\bibinfo {author} {\bibfnamefont {T.}~\bibnamefont
  {Jolicoeur}},\ }\href {\doibase 10.1103/PhysRevB.95.075201} {\bibfield
  {journal} {\bibinfo  {journal} {Phys. Rev. B}\ }\textbf {\bibinfo {volume}
  {95}},\ \bibinfo {pages} {075201} (\bibinfo {year} {2017})}\BibitemShut
  {NoStop}%
\bibitem [{\citenamefont {Liu}\ \emph {et~al.}(2018)\citenamefont {Liu},
  \citenamefont {Gromov},\ and\ \citenamefont {Papi\ifmmode~\acute{c}\else
  \'{c}\fi{}}}]{PapicMain}%
  \BibitemOpen
  \bibfield  {author} {\bibinfo {author} {\bibfnamefont {Z.}~\bibnamefont
  {Liu}}, \bibinfo {author} {\bibfnamefont {A.}~\bibnamefont {Gromov}}, \ and\
  \bibinfo {author} {\bibfnamefont {Z.}~\bibnamefont
  {Papi\ifmmode~\acute{c}\else \'{c}\fi{}}},\ }\href {\doibase
  10.1103/PhysRevB.98.155140} {\bibfield  {journal} {\bibinfo  {journal} {Phys.
  Rev. B}\ }\textbf {\bibinfo {volume} {98}},\ \bibinfo {pages} {155140}
  (\bibinfo {year} {2018})}\BibitemShut {NoStop}%
\bibitem [{\citenamefont {Lapa}\ \emph {et~al.}(2019)\citenamefont {Lapa},
  \citenamefont {Gromov},\ and\ \citenamefont {Hughes}}]{Lapa19}%
  \BibitemOpen
  \bibfield  {author} {\bibinfo {author} {\bibfnamefont {M.~F.}\ \bibnamefont
  {Lapa}}, \bibinfo {author} {\bibfnamefont {A.}~\bibnamefont {Gromov}}, \ and\
  \bibinfo {author} {\bibfnamefont {T.~L.}\ \bibnamefont {Hughes}},\ }\href
  {\doibase 10.1103/PhysRevB.99.075115} {\bibfield  {journal} {\bibinfo
  {journal} {Phys. Rev. B}\ }\textbf {\bibinfo {volume} {99}},\ \bibinfo
  {pages} {075115} (\bibinfo {year} {2019})}\BibitemShut {NoStop}%
\bibitem [{\citenamefont {Liou}\ \emph {et~al.}(2019)\citenamefont {Liou},
  \citenamefont {Haldane}, \citenamefont {Yang},\ and\ \citenamefont
  {Rezayi}}]{Liou19}%
  \BibitemOpen
  \bibfield  {author} {\bibinfo {author} {\bibfnamefont {S.-F.}\ \bibnamefont
  {Liou}}, \bibinfo {author} {\bibfnamefont {F.~D.~M.}\ \bibnamefont
  {Haldane}}, \bibinfo {author} {\bibfnamefont {K.}~\bibnamefont {Yang}}, \
  and\ \bibinfo {author} {\bibfnamefont {E.~H.}\ \bibnamefont {Rezayi}},\
  }\href {\doibase 10.1103/PhysRevLett.123.146801} {\bibfield  {journal}
  {\bibinfo  {journal} {Phys. Rev. Lett.}\ }\textbf {\bibinfo {volume} {123}},\
  \bibinfo {pages} {146801} (\bibinfo {year} {2019})}\BibitemShut {NoStop}%
\bibitem [{\citenamefont {Nguyen}\ and\ \citenamefont
  {Son}(2021)}]{Nguyen2021}%
  \BibitemOpen
  \bibfield  {author} {\bibinfo {author} {\bibfnamefont {D.~X.}\ \bibnamefont
  {Nguyen}}\ and\ \bibinfo {author} {\bibfnamefont {D.~T.}\ \bibnamefont
  {Son}},\ }\href {\doibase 10.1103/PhysRevResearch.3.023040} {\bibfield
  {journal} {\bibinfo  {journal} {Phys. Rev. Research}\ }\textbf {\bibinfo
  {volume} {3}},\ \bibinfo {pages} {023040} (\bibinfo {year}
  {2021})}\BibitemShut {NoStop}%
\bibitem [{\citenamefont {Wang}\ and\ \citenamefont {Yang}()}]{yuzhu}%
  \BibitemOpen
  \bibfield  {author} {\bibinfo {author} {\bibfnamefont {Y.}~\bibnamefont
  {Wang}}\ and\ \bibinfo {author} {\bibfnamefont {B.}~\bibnamefont {Yang}},\
  }\href@noop {} {\enquote {\bibinfo {title} {Analytic exposition of the
  graviton modes in fractional quantum hall effects and its physical
  implications},}\ }\Eprint {http://arxiv.org/abs/arXiv:2109.08816}
  {arXiv:arXiv:2109.08816} \BibitemShut {NoStop}%
\bibitem [{\citenamefont {Xia}\ \emph {et~al.}(2011)\citenamefont {Xia},
  \citenamefont {Eisenstein}, \citenamefont {Pfeiffer},\ and\ \citenamefont
  {West}}]{Xia2011}%
  \BibitemOpen
  \bibfield  {author} {\bibinfo {author} {\bibfnamefont {J.}~\bibnamefont
  {Xia}}, \bibinfo {author} {\bibfnamefont {J.~P.}\ \bibnamefont {Eisenstein}},
  \bibinfo {author} {\bibfnamefont {L.~N.}\ \bibnamefont {Pfeiffer}}, \ and\
  \bibinfo {author} {\bibfnamefont {K.~W.}\ \bibnamefont {West}},\ }\href
  {\doibase 10.1038/nphys2118} {\bibfield  {journal} {\bibinfo  {journal}
  {Nature Physics}\ }\textbf {\bibinfo {volume} {7}},\ \bibinfo {pages} {845}
  (\bibinfo {year} {2011})}\BibitemShut {NoStop}%
\bibitem [{\citenamefont {Samkharadze}\ \emph {et~al.}(2016)\citenamefont
  {Samkharadze}, \citenamefont {Schreiber}, \citenamefont {Gardner},
  \citenamefont {Manfra}, \citenamefont {Fradkin},\ and\ \citenamefont
  {Cs{\'a}thy}}]{Samkharadze2016}%
  \BibitemOpen
  \bibfield  {author} {\bibinfo {author} {\bibfnamefont {N.}~\bibnamefont
  {Samkharadze}}, \bibinfo {author} {\bibfnamefont {K.~A.}\ \bibnamefont
  {Schreiber}}, \bibinfo {author} {\bibfnamefont {G.~C.}\ \bibnamefont
  {Gardner}}, \bibinfo {author} {\bibfnamefont {M.~J.}\ \bibnamefont {Manfra}},
  \bibinfo {author} {\bibfnamefont {E.}~\bibnamefont {Fradkin}}, \ and\
  \bibinfo {author} {\bibfnamefont {G.~A.}\ \bibnamefont {Cs{\'a}thy}},\ }\href
  {\doibase 10.1038/nphys3523} {\bibfield  {journal} {\bibinfo  {journal}
  {Nature Physics}\ }\textbf {\bibinfo {volume} {12}},\ \bibinfo {pages} {191}
  (\bibinfo {year} {2016})}\BibitemShut {NoStop}%
\bibitem [{\citenamefont {Regnault}\ \emph {et~al.}(2017)\citenamefont
  {Regnault}, \citenamefont {Maciejko}, \citenamefont {Kivelson},\ and\
  \citenamefont {Sondhi}}]{PRBNem}%
  \BibitemOpen
  \bibfield  {author} {\bibinfo {author} {\bibfnamefont {N.}~\bibnamefont
  {Regnault}}, \bibinfo {author} {\bibfnamefont {J.}~\bibnamefont {Maciejko}},
  \bibinfo {author} {\bibfnamefont {S.~A.}\ \bibnamefont {Kivelson}}, \ and\
  \bibinfo {author} {\bibfnamefont {S.~L.}\ \bibnamefont {Sondhi}},\ }\href
  {\doibase 10.1103/PhysRevB.96.035150} {\bibfield  {journal} {\bibinfo
  {journal} {Phys. Rev. B}\ }\textbf {\bibinfo {volume} {96}},\ \bibinfo
  {pages} {035150} (\bibinfo {year} {2017})}\BibitemShut {NoStop}%
\bibitem [{\citenamefont {You}\ \emph {et~al.}(2014)\citenamefont {You},
  \citenamefont {Cho},\ and\ \citenamefont {Fradkin}}]{PRXNem}%
  \BibitemOpen
  \bibfield  {author} {\bibinfo {author} {\bibfnamefont {Y.}~\bibnamefont
  {You}}, \bibinfo {author} {\bibfnamefont {G.~Y.}\ \bibnamefont {Cho}}, \ and\
  \bibinfo {author} {\bibfnamefont {E.}~\bibnamefont {Fradkin}},\ }\href
  {\doibase 10.1103/PhysRevX.4.041050} {\bibfield  {journal} {\bibinfo
  {journal} {Phys. Rev. X}\ }\textbf {\bibinfo {volume} {4}},\ \bibinfo {pages}
  {041050} (\bibinfo {year} {2014})}\BibitemShut {NoStop}%
\bibitem [{\citenamefont {Yang}(2020)}]{PRRNem}%
  \BibitemOpen
  \bibfield  {author} {\bibinfo {author} {\bibfnamefont {B.}~\bibnamefont
  {Yang}},\ }\href {\doibase 10.1103/PhysRevResearch.2.033362} {\bibfield
  {journal} {\bibinfo  {journal} {Phys. Rev. Research}\ }\textbf {\bibinfo
  {volume} {2}},\ \bibinfo {pages} {033362} (\bibinfo {year}
  {2020})}\BibitemShut {NoStop}%
\bibitem [{\citenamefont {Zaletel}\ and\ \citenamefont
  {Mong}(2012)}]{Zaletel2012}%
  \BibitemOpen
  \bibfield  {author} {\bibinfo {author} {\bibfnamefont {M.~P.}\ \bibnamefont
  {Zaletel}}\ and\ \bibinfo {author} {\bibfnamefont {R.~S.~K.}\ \bibnamefont
  {Mong}},\ }\href {\doibase 10.1103/PhysRevB.86.245305} {\bibfield  {journal}
  {\bibinfo  {journal} {Phys. Rev. B}\ }\textbf {\bibinfo {volume} {86}},\
  \bibinfo {pages} {245305} (\bibinfo {year} {2012})}\BibitemShut {NoStop}%
\bibitem [{\citenamefont {Estienne}\ \emph {et~al.}(2013)\citenamefont
  {Estienne}, \citenamefont {Papi\ifmmode~\acute{c}\else \'{c}\fi{}},
  \citenamefont {Regnault},\ and\ \citenamefont {Bernevig}}]{Estienne2013}%
  \BibitemOpen
  \bibfield  {author} {\bibinfo {author} {\bibfnamefont {B.}~\bibnamefont
  {Estienne}}, \bibinfo {author} {\bibfnamefont {Z.}~\bibnamefont
  {Papi\ifmmode~\acute{c}\else \'{c}\fi{}}}, \bibinfo {author} {\bibfnamefont
  {N.}~\bibnamefont {Regnault}}, \ and\ \bibinfo {author} {\bibfnamefont
  {B.~A.}\ \bibnamefont {Bernevig}},\ }\href {\doibase
  10.1103/PhysRevB.87.161112} {\bibfield  {journal} {\bibinfo  {journal} {Phys.
  Rev. B}\ }\textbf {\bibinfo {volume} {87}},\ \bibinfo {pages} {161112}
  (\bibinfo {year} {2013})}\BibitemShut {NoStop}%
\bibitem [{\citenamefont {Rezayi}\ and\ \citenamefont
  {Haldane}(1994)}]{HaldaneCyl}%
  \BibitemOpen
  \bibfield  {author} {\bibinfo {author} {\bibfnamefont {E.~H.}\ \bibnamefont
  {Rezayi}}\ and\ \bibinfo {author} {\bibfnamefont {F.~D.~M.}\ \bibnamefont
  {Haldane}},\ }\href {\doibase 10.1103/PhysRevB.50.17199} {\bibfield
  {journal} {\bibinfo  {journal} {Phys. Rev. B}\ }\textbf {\bibinfo {volume}
  {50}},\ \bibinfo {pages} {17199} (\bibinfo {year} {1994})}\BibitemShut
  {NoStop}%
\bibitem [{\citenamefont {Bergholtz}\ and\ \citenamefont
  {Karlhede}(2005)}]{Bergholtz2005}%
  \BibitemOpen
  \bibfield  {author} {\bibinfo {author} {\bibfnamefont {E.~J.}\ \bibnamefont
  {Bergholtz}}\ and\ \bibinfo {author} {\bibfnamefont {A.}~\bibnamefont
  {Karlhede}},\ }\href {\doibase 10.1103/PhysRevLett.94.026802} {\bibfield
  {journal} {\bibinfo  {journal} {Phys. Rev. Lett.}\ }\textbf {\bibinfo
  {volume} {94}},\ \bibinfo {pages} {026802} (\bibinfo {year}
  {2005})}\BibitemShut {NoStop}%
\bibitem [{\citenamefont {Seidel}\ \emph {et~al.}(2005)\citenamefont {Seidel},
  \citenamefont {Fu}, \citenamefont {Lee}, \citenamefont {Leinaas},\ and\
  \citenamefont {Moore}}]{Seidel2005}%
  \BibitemOpen
  \bibfield  {author} {\bibinfo {author} {\bibfnamefont {A.}~\bibnamefont
  {Seidel}}, \bibinfo {author} {\bibfnamefont {H.}~\bibnamefont {Fu}}, \bibinfo
  {author} {\bibfnamefont {D.-H.}\ \bibnamefont {Lee}}, \bibinfo {author}
  {\bibfnamefont {J.~M.}\ \bibnamefont {Leinaas}}, \ and\ \bibinfo {author}
  {\bibfnamefont {J.}~\bibnamefont {Moore}},\ }\href {\doibase
  10.1103/PhysRevLett.95.266405} {\bibfield  {journal} {\bibinfo  {journal}
  {Phys. Rev. Lett.}\ }\textbf {\bibinfo {volume} {95}},\ \bibinfo {pages}
  {266405} (\bibinfo {year} {2005})}\BibitemShut {NoStop}%
\bibitem [{\citenamefont {Bergholtz}\ and\ \citenamefont
  {Karlhede}(2008)}]{Bergholtz2008}%
  \BibitemOpen
  \bibfield  {author} {\bibinfo {author} {\bibfnamefont {E.~J.}\ \bibnamefont
  {Bergholtz}}\ and\ \bibinfo {author} {\bibfnamefont {A.}~\bibnamefont
  {Karlhede}},\ }\href {\doibase 10.1103/PhysRevB.77.155308} {\bibfield
  {journal} {\bibinfo  {journal} {Phys. Rev. B}\ }\textbf {\bibinfo {volume}
  {77}},\ \bibinfo {pages} {155308} (\bibinfo {year} {2008})}\BibitemShut
  {NoStop}%
\bibitem [{\citenamefont {Rahmani}\ \emph {et~al.}(2020)\citenamefont
  {Rahmani}, \citenamefont {Sung}, \citenamefont {Putterman}, \citenamefont
  {Roushan}, \citenamefont {Ghaemi},\ and\ \citenamefont {Jiang}}]{Rahmani}%
  \BibitemOpen
  \bibfield  {author} {\bibinfo {author} {\bibfnamefont {A.}~\bibnamefont
  {Rahmani}}, \bibinfo {author} {\bibfnamefont {K.~J.}\ \bibnamefont {Sung}},
  \bibinfo {author} {\bibfnamefont {H.}~\bibnamefont {Putterman}}, \bibinfo
  {author} {\bibfnamefont {P.}~\bibnamefont {Roushan}}, \bibinfo {author}
  {\bibfnamefont {P.}~\bibnamefont {Ghaemi}}, \ and\ \bibinfo {author}
  {\bibfnamefont {Z.}~\bibnamefont {Jiang}},\ }\href {\doibase
  10.1103/PRXQuantum.1.020309} {\bibfield  {journal} {\bibinfo  {journal} {PRX
  Quantum}\ }\textbf {\bibinfo {volume} {1}},\ \bibinfo {pages} {020309}
  (\bibinfo {year} {2020})}\BibitemShut {NoStop}%
\bibitem [{\citenamefont {Nakamura}\ \emph {et~al.}(2012)\citenamefont
  {Nakamura}, \citenamefont {Wang},\ and\ \citenamefont
  {Bergholtz}}]{Nakamura2012}%
  \BibitemOpen
  \bibfield  {author} {\bibinfo {author} {\bibfnamefont {M.}~\bibnamefont
  {Nakamura}}, \bibinfo {author} {\bibfnamefont {Z.-Y.}\ \bibnamefont {Wang}},
  \ and\ \bibinfo {author} {\bibfnamefont {E.~J.}\ \bibnamefont {Bergholtz}},\
  }\href {\doibase 10.1103/PhysRevLett.109.016401} {\bibfield  {journal}
  {\bibinfo  {journal} {Phys. Rev. Lett.}\ }\textbf {\bibinfo {volume} {109}},\
  \bibinfo {pages} {016401} (\bibinfo {year} {2012})}\BibitemShut {NoStop}%
\bibitem [{\citenamefont {Wang}\ and\ \citenamefont
  {Nakamura}(2013)}]{Wang2013}%
  \BibitemOpen
  \bibfield  {author} {\bibinfo {author} {\bibfnamefont {Z.-Y.}\ \bibnamefont
  {Wang}}\ and\ \bibinfo {author} {\bibfnamefont {M.}~\bibnamefont
  {Nakamura}},\ }\href {\doibase 10.1103/PhysRevB.87.245119} {\bibfield
  {journal} {\bibinfo  {journal} {Phys. Rev. B}\ }\textbf {\bibinfo {volume}
  {87}},\ \bibinfo {pages} {245119} (\bibinfo {year} {2013})}\BibitemShut
  {NoStop}%
\bibitem [{\citenamefont {Soul\'e}\ and\ \citenamefont
  {Jolicoeur}(2012)}]{Jolicoeur2012}%
  \BibitemOpen
  \bibfield  {author} {\bibinfo {author} {\bibfnamefont {P.}~\bibnamefont
  {Soul\'e}}\ and\ \bibinfo {author} {\bibfnamefont {T.}~\bibnamefont
  {Jolicoeur}},\ }\href {\doibase 10.1103/PhysRevB.85.155116} {\bibfield
  {journal} {\bibinfo  {journal} {Phys. Rev. B}\ }\textbf {\bibinfo {volume}
  {85}},\ \bibinfo {pages} {155116} (\bibinfo {year} {2012})}\BibitemShut
  {NoStop}%
\bibitem [{\citenamefont {Saravanan}\ and\ \citenamefont
  {Saeed}(2022)}]{saravanan2022pauli}%
  \BibitemOpen
  \bibfield  {author} {\bibinfo {author} {\bibfnamefont {V.}~\bibnamefont
  {Saravanan}}\ and\ \bibinfo {author} {\bibfnamefont {S.~M.}\ \bibnamefont
  {Saeed}},\ }\href@noop {} {\enquote {\bibinfo {title} {Pauli error
  propagation-based gate reschedulingfor quantum circuit error mitigation},}\ }
  (\bibinfo {year} {2022}),\ \Eprint {http://arxiv.org/abs/2201.12946}
  {arXiv:2201.12946 [quant-ph]} \BibitemShut {NoStop}%
\bibitem [{\citenamefont {Abraham}\ and\ \citenamefont
  {et~al.}(2019)}]{Qiskit}%
  \BibitemOpen
  \bibfield  {author} {\bibinfo {author} {\bibfnamefont {H.}~\bibnamefont
  {Abraham}}\ and\ \bibinfo {author} {\bibnamefont {et~al.}},\ }\href {\doibase
  10.5281/zenodo.2562110} {\enquote {\bibinfo {title} {Qiskit: An open-source
  framework for quantum computing},}\ } (\bibinfo {year} {2019})\BibitemShut
  {NoStop}%
\bibitem [{\citenamefont {Davis}\ \emph {et~al.}(2020)\citenamefont {Davis},
  \citenamefont {Smith}, \citenamefont {Tudor}, \citenamefont {Sen},
  \citenamefont {Siddiqi},\ and\ \citenamefont {Iancu}}]{9259942}%
  \BibitemOpen
  \bibfield  {author} {\bibinfo {author} {\bibfnamefont {M.~G.}\ \bibnamefont
  {Davis}}, \bibinfo {author} {\bibfnamefont {E.}~\bibnamefont {Smith}},
  \bibinfo {author} {\bibfnamefont {A.}~\bibnamefont {Tudor}}, \bibinfo
  {author} {\bibfnamefont {K.}~\bibnamefont {Sen}}, \bibinfo {author}
  {\bibfnamefont {I.}~\bibnamefont {Siddiqi}}, \ and\ \bibinfo {author}
  {\bibfnamefont {C.}~\bibnamefont {Iancu}},\ }in\ \href {\doibase
  10.1109/QCE49297.2020.00036} {\emph {\bibinfo {booktitle} {2020 IEEE
  International Conference on Quantum Computing and Engineering (QCE)}}}\
  (\bibinfo {year} {2020})\ pp.\ \bibinfo {pages} {223--234}\BibitemShut
  {NoStop}%
\bibitem [{\citenamefont {Werschnik}\ and\ \citenamefont
  {Gross}(2007)}]{werschnik_quantum_2007}%
  \BibitemOpen
  \bibfield  {author} {\bibinfo {author} {\bibfnamefont {J.}~\bibnamefont
  {Werschnik}}\ and\ \bibinfo {author} {\bibfnamefont {E.~K.~U.}\ \bibnamefont
  {Gross}},\ }\href {http://arxiv.org/abs/0707.1883} {\bibfield  {journal}
  {\bibinfo  {journal} {arXiv:0707.1883}\ } (\bibinfo {year}
  {2007})}\BibitemShut {NoStop}%
\bibitem [{\citenamefont {Petersen}\ and\ \citenamefont
  {Dong}(2010)}]{petersen_quantum_2010}%
  \BibitemOpen
  \bibfield  {author} {\bibinfo {author} {\bibfnamefont {I.}~\bibnamefont
  {Petersen}}\ and\ \bibinfo {author} {\bibfnamefont {D.}~\bibnamefont
  {Dong}},\ }\href {\doibase 10.1049/iet-cta.2009.0508} {\bibfield  {journal}
  {\bibinfo  {journal} {IET Control Theory \& Applications}\ }\textbf {\bibinfo
  {volume} {4}},\ \bibinfo {pages} {2651} (\bibinfo {year} {2010})}\BibitemShut
  {NoStop}%
\bibitem [{\citenamefont {Rahmani}(2013)}]{Rahmani13}%
  \BibitemOpen
  \bibfield  {author} {\bibinfo {author} {\bibfnamefont {A.}~\bibnamefont
  {Rahmani}},\ }\href {https://doi.org/10.1142/S0217984913300196} {\bibfield
  {journal} {\bibinfo  {journal} {Mod. Phys. Lett. B}\ }\textbf {\bibinfo
  {volume} {27}},\ \bibinfo {pages} {1330019} (\bibinfo {year}
  {2013})}\BibitemShut {NoStop}%
\bibitem [{\citenamefont {Peruzzo}\ \emph {et~al.}(2014)\citenamefont
  {Peruzzo}, \citenamefont {McClean}, \citenamefont {Shadbolt}, \citenamefont
  {Yung}, \citenamefont {Zhou}, \citenamefont {Love}, \citenamefont
  {Aspuru-Guzik},\ and\ \citenamefont {O'Brien}}]{Peruzzo:14}%
  \BibitemOpen
  \bibfield  {author} {\bibinfo {author} {\bibfnamefont {A.}~\bibnamefont
  {Peruzzo}}, \bibinfo {author} {\bibfnamefont {J.}~\bibnamefont {McClean}},
  \bibinfo {author} {\bibfnamefont {P.}~\bibnamefont {Shadbolt}}, \bibinfo
  {author} {\bibfnamefont {M.~H.}\ \bibnamefont {Yung}}, \bibinfo {author}
  {\bibfnamefont {X.~Q.}\ \bibnamefont {Zhou}}, \bibinfo {author}
  {\bibfnamefont {P.~J.}\ \bibnamefont {Love}}, \bibinfo {author}
  {\bibfnamefont {A.}~\bibnamefont {Aspuru-Guzik}}, \ and\ \bibinfo {author}
  {\bibfnamefont {J.~L.}\ \bibnamefont {O'Brien}},\ }\href
  {https://www.nature.com/articles/ncomms5213} {\bibfield  {journal} {\bibinfo
  {journal} {Nat. Comm.}\ }\textbf {\bibinfo {volume} {5}},\ \bibinfo {pages}
  {4213} (\bibinfo {year} {2014})}\BibitemShut {NoStop}%
\bibitem [{\citenamefont {Wecker}\ \emph {et~al.}(2015)\citenamefont {Wecker},
  \citenamefont {Hastings},\ and\ \citenamefont {Troyer}}]{Wecker:15}%
  \BibitemOpen
  \bibfield  {author} {\bibinfo {author} {\bibfnamefont {D.}~\bibnamefont
  {Wecker}}, \bibinfo {author} {\bibfnamefont {M.~B.}\ \bibnamefont
  {Hastings}}, \ and\ \bibinfo {author} {\bibfnamefont {M.}~\bibnamefont
  {Troyer}},\ }\href {\doibase 10.1103/PhysRevA.92.042303} {\bibfield
  {journal} {\bibinfo  {journal} {Phys. Rev. A}\ }\textbf {\bibinfo {volume}
  {92}},\ \bibinfo {pages} {042303} (\bibinfo {year} {2015})}\BibitemShut
  {NoStop}%
\bibitem [{\citenamefont {Wecker}\ \emph {et~al.}(2016)\citenamefont {Wecker},
  \citenamefont {Hastings},\ and\ \citenamefont {Troyer}}]{wecker}%
  \BibitemOpen
  \bibfield  {author} {\bibinfo {author} {\bibfnamefont {D.}~\bibnamefont
  {Wecker}}, \bibinfo {author} {\bibfnamefont {M.~B.}\ \bibnamefont
  {Hastings}}, \ and\ \bibinfo {author} {\bibfnamefont {M.}~\bibnamefont
  {Troyer}},\ }\href {\doibase 10.1103/PhysRevA.94.022309} {\bibfield
  {journal} {\bibinfo  {journal} {Phys. Rev. A}\ }\textbf {\bibinfo {volume}
  {94}},\ \bibinfo {pages} {022309} (\bibinfo {year} {2016})}\BibitemShut
  {NoStop}%
\bibitem [{\citenamefont {McClean}\ \emph {et~al.}(2016)\citenamefont
  {McClean}, \citenamefont {Romero}, \citenamefont {Babbush},\ and\
  \citenamefont {Aspuru-Guzik}}]{McClean_2016}%
  \BibitemOpen
  \bibfield  {author} {\bibinfo {author} {\bibfnamefont {J.~R.}\ \bibnamefont
  {McClean}}, \bibinfo {author} {\bibfnamefont {J.}~\bibnamefont {Romero}},
  \bibinfo {author} {\bibfnamefont {R.}~\bibnamefont {Babbush}}, \ and\
  \bibinfo {author} {\bibfnamefont {A.}~\bibnamefont {Aspuru-Guzik}},\ }\href
  {\doibase 10.1088/1367-2630/18/2/023023} {\bibfield  {journal} {\bibinfo
  {journal} {New Journal of Physics}\ }\textbf {\bibinfo {volume} {18}},\
  \bibinfo {pages} {023023} (\bibinfo {year} {2016})}\BibitemShut {NoStop}%
\bibitem [{\citenamefont {Farhi}\ \emph {et~al.}()\citenamefont {Farhi},
  \citenamefont {Goldstone},\ and\ \citenamefont {Gutmann}}]{Farhi}%
  \BibitemOpen
  \bibfield  {author} {\bibinfo {author} {\bibfnamefont {E.}~\bibnamefont
  {Farhi}}, \bibinfo {author} {\bibfnamefont {J.}~\bibnamefont {Goldstone}}, \
  and\ \bibinfo {author} {\bibfnamefont {S.}~\bibnamefont {Gutmann}},\
  }\href@noop {} {\enquote {\bibinfo {title} {{A Quantum Approximate
  Optimization Algorithm}},}\ }\Eprint {http://arxiv.org/abs/1411.4028}
  {arXiv:1411.4028} \BibitemShut {NoStop}%
\bibitem [{\citenamefont {Farhi}\ and\ \citenamefont {Harrow}()}]{Farhi:3}%
  \BibitemOpen
  \bibfield  {author} {\bibinfo {author} {\bibfnamefont {E.}~\bibnamefont
  {Farhi}}\ and\ \bibinfo {author} {\bibfnamefont {A.~W.}\ \bibnamefont
  {Harrow}},\ }\href@noop {} {\enquote {\bibinfo {title} {Quantum supremacy
  through the quantum approximate optimization algorithm},}\ }\Eprint
  {http://arxiv.org/abs/1602.07674} {arXiv:1602.07674} \BibitemShut {NoStop}%
\bibitem [{\citenamefont {Yang}\ \emph {et~al.}(2017)\citenamefont {Yang},
  \citenamefont {Rahmani}, \citenamefont {Shabani}, \citenamefont {Neven},\
  and\ \citenamefont {Chamon}}]{Yang17}%
  \BibitemOpen
  \bibfield  {author} {\bibinfo {author} {\bibfnamefont {Z.-C.}\ \bibnamefont
  {Yang}}, \bibinfo {author} {\bibfnamefont {A.}~\bibnamefont {Rahmani}},
  \bibinfo {author} {\bibfnamefont {A.}~\bibnamefont {Shabani}}, \bibinfo
  {author} {\bibfnamefont {H.}~\bibnamefont {Neven}}, \ and\ \bibinfo {author}
  {\bibfnamefont {C.}~\bibnamefont {Chamon}},\ }\href {\doibase
  10.1103/PhysRevX.7.021027} {\bibfield  {journal} {\bibinfo  {journal} {Phys.
  Rev. X}\ }\textbf {\bibinfo {volume} {7}},\ \bibinfo {pages} {021027}
  (\bibinfo {year} {2017})}\BibitemShut {NoStop}%
\bibitem [{\citenamefont {Wang}\ \emph {et~al.}(2018)\citenamefont {Wang},
  \citenamefont {Hadfield}, \citenamefont {Jiang},\ and\ \citenamefont
  {Rieffel}}]{Wang18}%
  \BibitemOpen
  \bibfield  {author} {\bibinfo {author} {\bibfnamefont {Z.}~\bibnamefont
  {Wang}}, \bibinfo {author} {\bibfnamefont {S.}~\bibnamefont {Hadfield}},
  \bibinfo {author} {\bibfnamefont {Z.}~\bibnamefont {Jiang}}, \ and\ \bibinfo
  {author} {\bibfnamefont {E.~G.}\ \bibnamefont {Rieffel}},\ }\href {\doibase
  10.1103/PhysRevA.97.022304} {\bibfield  {journal} {\bibinfo  {journal} {Phys.
  Rev. A}\ }\textbf {\bibinfo {volume} {97}},\ \bibinfo {pages} {022304}
  (\bibinfo {year} {2018})}\BibitemShut {NoStop}%
\bibitem [{\citenamefont {Zhou}\ \emph {et~al.}(2020)\citenamefont {Zhou},
  \citenamefont {Wang}, \citenamefont {Choi}, \citenamefont {Pichler},\ and\
  \citenamefont {Lukin}}]{Zhou20}%
  \BibitemOpen
  \bibfield  {author} {\bibinfo {author} {\bibfnamefont {L.}~\bibnamefont
  {Zhou}}, \bibinfo {author} {\bibfnamefont {S.-T.}\ \bibnamefont {Wang}},
  \bibinfo {author} {\bibfnamefont {S.}~\bibnamefont {Choi}}, \bibinfo {author}
  {\bibfnamefont {H.}~\bibnamefont {Pichler}}, \ and\ \bibinfo {author}
  {\bibfnamefont {M.~D.}\ \bibnamefont {Lukin}},\ }\href {\doibase
  10.1103/PhysRevX.10.021067} {\bibfield  {journal} {\bibinfo  {journal} {Phys.
  Rev. X}\ }\textbf {\bibinfo {volume} {10}},\ \bibinfo {pages} {021067}
  (\bibinfo {year} {2020})}\BibitemShut {NoStop}%
\bibitem [{\citenamefont {Lin}\ \emph {et~al.}(2021)\citenamefont {Lin},
  \citenamefont {Dilip}, \citenamefont {Green}, \citenamefont {Smith},\ and\
  \citenamefont {Pollmann}}]{green1}%
  \BibitemOpen
  \bibfield  {author} {\bibinfo {author} {\bibfnamefont {S.-H.}\ \bibnamefont
  {Lin}}, \bibinfo {author} {\bibfnamefont {R.}~\bibnamefont {Dilip}}, \bibinfo
  {author} {\bibfnamefont {A.~G.}\ \bibnamefont {Green}}, \bibinfo {author}
  {\bibfnamefont {A.}~\bibnamefont {Smith}}, \ and\ \bibinfo {author}
  {\bibfnamefont {F.}~\bibnamefont {Pollmann}},\ }\href {\doibase
  10.1103/PRXQuantum.2.010342} {\bibfield  {journal} {\bibinfo  {journal} {PRX
  Quantum}\ }\textbf {\bibinfo {volume} {2}},\ \bibinfo {pages} {010342}
  (\bibinfo {year} {2021})}\BibitemShut {NoStop}%
\bibitem [{\citenamefont {Kokail}\ \emph {et~al.}(2019)\citenamefont {Kokail},
  \citenamefont {Maier}, \citenamefont {van Bijnen}, \citenamefont {Brydges},
  \citenamefont {Joshi}, \citenamefont {Jurcevic}, \citenamefont {Muschik},
  \citenamefont {Silvi}, \citenamefont {Blatt}, \citenamefont {Roos},\ and\
  \citenamefont {Zoller}}]{VQAnat1}%
  \BibitemOpen
  \bibfield  {author} {\bibinfo {author} {\bibfnamefont {C.}~\bibnamefont
  {Kokail}}, \bibinfo {author} {\bibfnamefont {C.}~\bibnamefont {Maier}},
  \bibinfo {author} {\bibfnamefont {R.}~\bibnamefont {van Bijnen}}, \bibinfo
  {author} {\bibfnamefont {T.}~\bibnamefont {Brydges}}, \bibinfo {author}
  {\bibfnamefont {M.~K.}\ \bibnamefont {Joshi}}, \bibinfo {author}
  {\bibfnamefont {P.}~\bibnamefont {Jurcevic}}, \bibinfo {author}
  {\bibfnamefont {C.~A.}\ \bibnamefont {Muschik}}, \bibinfo {author}
  {\bibfnamefont {P.}~\bibnamefont {Silvi}}, \bibinfo {author} {\bibfnamefont
  {R.}~\bibnamefont {Blatt}}, \bibinfo {author} {\bibfnamefont {C.~F.}\
  \bibnamefont {Roos}}, \ and\ \bibinfo {author} {\bibfnamefont
  {P.}~\bibnamefont {Zoller}},\ }\href {\doibase 10.1038/s41586-019-1177-4}
  {\bibfield  {journal} {\bibinfo  {journal} {Nature}\ }\textbf {\bibinfo
  {volume} {569}},\ \bibinfo {pages} {355} (\bibinfo {year}
  {2019})}\BibitemShut {NoStop}%
\bibitem [{\citenamefont {Kandala}\ \emph {et~al.}(2017)\citenamefont
  {Kandala}, \citenamefont {Mezzacapo}, \citenamefont {Temme}, \citenamefont
  {Takita}, \citenamefont {Brink}, \citenamefont {Chow},\ and\ \citenamefont
  {Gambetta}}]{VQAnat2}%
  \BibitemOpen
  \bibfield  {author} {\bibinfo {author} {\bibfnamefont {A.}~\bibnamefont
  {Kandala}}, \bibinfo {author} {\bibfnamefont {A.}~\bibnamefont {Mezzacapo}},
  \bibinfo {author} {\bibfnamefont {K.}~\bibnamefont {Temme}}, \bibinfo
  {author} {\bibfnamefont {M.}~\bibnamefont {Takita}}, \bibinfo {author}
  {\bibfnamefont {M.}~\bibnamefont {Brink}}, \bibinfo {author} {\bibfnamefont
  {J.~M.}\ \bibnamefont {Chow}}, \ and\ \bibinfo {author} {\bibfnamefont
  {J.~M.}\ \bibnamefont {Gambetta}},\ }\href {\doibase 10.1038/nature23879}
  {\bibfield  {journal} {\bibinfo  {journal} {Nature}\ }\textbf {\bibinfo
  {volume} {549}},\ \bibinfo {pages} {242} (\bibinfo {year}
  {2017})}\BibitemShut {NoStop}%
\bibitem [{\citenamefont {Maciejko}\ \emph {et~al.}(2013)\citenamefont
  {Maciejko}, \citenamefont {Hsu}, \citenamefont {Kivelson}, \citenamefont
  {Park},\ and\ \citenamefont {Sondhi}}]{maciejko2013field}%
  \BibitemOpen
  \bibfield  {author} {\bibinfo {author} {\bibfnamefont {J.}~\bibnamefont
  {Maciejko}}, \bibinfo {author} {\bibfnamefont {B.}~\bibnamefont {Hsu}},
  \bibinfo {author} {\bibfnamefont {S.~A.}\ \bibnamefont {Kivelson}}, \bibinfo
  {author} {\bibfnamefont {Y.}~\bibnamefont {Park}}, \ and\ \bibinfo {author}
  {\bibfnamefont {S.~L.}\ \bibnamefont {Sondhi}},\ }\href {\doibase
  10.1103/PhysRevB.88.125137} {\bibfield  {journal} {\bibinfo  {journal} {Phys.
  Rev. B}\ }\textbf {\bibinfo {volume} {88}},\ \bibinfo {pages} {125137}
  (\bibinfo {year} {2013})}\BibitemShut {NoStop}%
\bibitem [{\citenamefont {Bernevig}\ and\ \citenamefont
  {Haldane}(2008)}]{Bernevig2008}%
  \BibitemOpen
  \bibfield  {author} {\bibinfo {author} {\bibfnamefont {B.~A.}\ \bibnamefont
  {Bernevig}}\ and\ \bibinfo {author} {\bibfnamefont {F.~D.~M.}\ \bibnamefont
  {Haldane}},\ }\href {\doibase 10.1103/PhysRevLett.100.246802} {\bibfield
  {journal} {\bibinfo  {journal} {Phys. Rev. Lett.}\ }\textbf {\bibinfo
  {volume} {100}},\ \bibinfo {pages} {246802} (\bibinfo {year}
  {2008})}\BibitemShut {NoStop}%
\bibitem [{SOM(2021)}]{SOM}%
  \BibitemOpen
  \href@noop {} {\enquote {\bibinfo {title} {Supplemental online material},}\ }
  (\bibinfo {year} {2021})\BibitemShut {NoStop}%
\bibitem [{\citenamefont {Kitaev}\ and\ \citenamefont
  {Preskill}(2006)}]{KitaevPreskill}%
  \BibitemOpen
  \bibfield  {author} {\bibinfo {author} {\bibfnamefont {A.}~\bibnamefont
  {Kitaev}}\ and\ \bibinfo {author} {\bibfnamefont {J.}~\bibnamefont
  {Preskill}},\ }\href {\doibase 10.1103/PhysRevLett.96.110404} {\bibfield
  {journal} {\bibinfo  {journal} {Phys. Rev. Lett.}\ }\textbf {\bibinfo
  {volume} {96}},\ \bibinfo {pages} {110404} (\bibinfo {year}
  {2006})}\BibitemShut {NoStop}%
\bibitem [{\citenamefont {Levin}\ and\ \citenamefont {Wen}(2006)}]{LevinWen}%
  \BibitemOpen
  \bibfield  {author} {\bibinfo {author} {\bibfnamefont {M.}~\bibnamefont
  {Levin}}\ and\ \bibinfo {author} {\bibfnamefont {X.-G.}\ \bibnamefont
  {Wen}},\ }\href {\doibase 10.1103/PhysRevLett.96.110405} {\bibfield
  {journal} {\bibinfo  {journal} {Phys. Rev. Lett.}\ }\textbf {\bibinfo
  {volume} {96}},\ \bibinfo {pages} {110405} (\bibinfo {year}
  {2006})}\BibitemShut {NoStop}%
\bibitem [{\citenamefont {Yang}\ \emph
  {et~al.}(2012{\natexlab{b}})\citenamefont {Yang}, \citenamefont
  {Papi\ifmmode~\acute{c}\else \'{c}\fi{}}, \citenamefont {Rezayi},
  \citenamefont {Bhatt},\ and\ \citenamefont
  {Haldane}}]{BoYangPhysRevB.85.165318}%
  \BibitemOpen
  \bibfield  {author} {\bibinfo {author} {\bibfnamefont {B.}~\bibnamefont
  {Yang}}, \bibinfo {author} {\bibfnamefont {Z.}~\bibnamefont
  {Papi\ifmmode~\acute{c}\else \'{c}\fi{}}}, \bibinfo {author} {\bibfnamefont
  {E.~H.}\ \bibnamefont {Rezayi}}, \bibinfo {author} {\bibfnamefont {R.~N.}\
  \bibnamefont {Bhatt}}, \ and\ \bibinfo {author} {\bibfnamefont {F.~D.~M.}\
  \bibnamefont {Haldane}},\ }\href {\doibase 10.1103/PhysRevB.85.165318}
  {\bibfield  {journal} {\bibinfo  {journal} {Phys. Rev. B}\ }\textbf {\bibinfo
  {volume} {85}},\ \bibinfo {pages} {165318} (\bibinfo {year}
  {2012}{\natexlab{b}})}\BibitemShut {NoStop}%
\bibitem [{\citenamefont {Gromov}\ and\ \citenamefont {Son}(2017)}]{GromovSon}%
  \BibitemOpen
  \bibfield  {author} {\bibinfo {author} {\bibfnamefont {A.}~\bibnamefont
  {Gromov}}\ and\ \bibinfo {author} {\bibfnamefont {D.~T.}\ \bibnamefont
  {Son}},\ }\href {\doibase 10.1103/PhysRevX.7.041032} {\bibfield  {journal}
  {\bibinfo  {journal} {Phys. Rev. X}\ }\textbf {\bibinfo {volume} {7}},\
  \bibinfo {pages} {041032} (\bibinfo {year} {2017})}\BibitemShut {NoStop}%
\bibitem [{\citenamefont {Yang}(2016)}]{KunYang}%
  \BibitemOpen
  \bibfield  {author} {\bibinfo {author} {\bibfnamefont {K.}~\bibnamefont
  {Yang}},\ }\href {\doibase 10.1103/PhysRevB.93.161302} {\bibfield  {journal}
  {\bibinfo  {journal} {Phys. Rev. B}\ }\textbf {\bibinfo {volume} {93}},\
  \bibinfo {pages} {161302} (\bibinfo {year} {2016})}\BibitemShut {NoStop}%
\bibitem [{\citenamefont {Moudgalya}\ \emph {et~al.}(2020)\citenamefont
  {Moudgalya}, \citenamefont {Bernevig},\ and\ \citenamefont
  {Regnault}}]{MoudgalyaThinTorus}%
  \BibitemOpen
  \bibfield  {author} {\bibinfo {author} {\bibfnamefont {S.}~\bibnamefont
  {Moudgalya}}, \bibinfo {author} {\bibfnamefont {B.~A.}\ \bibnamefont
  {Bernevig}}, \ and\ \bibinfo {author} {\bibfnamefont {N.}~\bibnamefont
  {Regnault}},\ }\href {\doibase 10.1103/PhysRevB.102.195150} {\bibfield
  {journal} {\bibinfo  {journal} {Phys. Rev. B}\ }\textbf {\bibinfo {volume}
  {102}},\ \bibinfo {pages} {195150} (\bibinfo {year} {2020})}\BibitemShut
  {NoStop}%
\bibitem [{\citenamefont {Moudgalya}\ \emph {et~al.}(2019)\citenamefont
  {Moudgalya}, \citenamefont {Prem}, \citenamefont {Nandkishore}, \citenamefont
  {Regnault},\ and\ \citenamefont {Bernevig}}]{MoudgalyaKrylov}%
  \BibitemOpen
  \bibfield  {author} {\bibinfo {author} {\bibfnamefont {S.}~\bibnamefont
  {Moudgalya}}, \bibinfo {author} {\bibfnamefont {A.}~\bibnamefont {Prem}},
  \bibinfo {author} {\bibfnamefont {R.}~\bibnamefont {Nandkishore}}, \bibinfo
  {author} {\bibfnamefont {N.}~\bibnamefont {Regnault}}, \ and\ \bibinfo
  {author} {\bibfnamefont {B.~A.}\ \bibnamefont {Bernevig}},\ }\href@noop {}
  {\enquote {\bibinfo {title} {Thermalization and its absence within krylov
  subspaces of a constrained hamiltonian},}\ } (\bibinfo {year} {2019}),\
  \Eprint {http://arxiv.org/abs/1910.14048} {arXiv:1910.14048
  [cond-mat.str-el]} \BibitemShut {NoStop}%
\bibitem [{\citenamefont {Feiguin}\ \emph {et~al.}(2007)\citenamefont
  {Feiguin}, \citenamefont {Trebst}, \citenamefont {Ludwig}, \citenamefont
  {Troyer}, \citenamefont {Kitaev}, \citenamefont {Wang},\ and\ \citenamefont
  {Freedman}}]{Feiguin2007}%
  \BibitemOpen
  \bibfield  {author} {\bibinfo {author} {\bibfnamefont {A.}~\bibnamefont
  {Feiguin}}, \bibinfo {author} {\bibfnamefont {S.}~\bibnamefont {Trebst}},
  \bibinfo {author} {\bibfnamefont {A.~W.~W.}\ \bibnamefont {Ludwig}}, \bibinfo
  {author} {\bibfnamefont {M.}~\bibnamefont {Troyer}}, \bibinfo {author}
  {\bibfnamefont {A.}~\bibnamefont {Kitaev}}, \bibinfo {author} {\bibfnamefont
  {Z.}~\bibnamefont {Wang}}, \ and\ \bibinfo {author} {\bibfnamefont {M.~H.}\
  \bibnamefont {Freedman}},\ }\href {\doibase 10.1103/PhysRevLett.98.160409}
  {\bibfield  {journal} {\bibinfo  {journal} {Phys. Rev. Lett.}\ }\textbf
  {\bibinfo {volume} {98}},\ \bibinfo {pages} {160409} (\bibinfo {year}
  {2007})}\BibitemShut {NoStop}%
\bibitem [{\citenamefont {Vidal}(2004)}]{VidalTEBD}%
  \BibitemOpen
  \bibfield  {author} {\bibinfo {author} {\bibfnamefont {G.}~\bibnamefont
  {Vidal}},\ }\href {\doibase 10.1103/PhysRevLett.93.040502} {\bibfield
  {journal} {\bibinfo  {journal} {Phys. Rev. Lett.}\ }\textbf {\bibinfo
  {volume} {93}},\ \bibinfo {pages} {040502} (\bibinfo {year}
  {2004})}\BibitemShut {NoStop}%
\bibitem [{\citenamefont {Haah}\ \emph {et~al.}(0)\citenamefont {Haah},
  \citenamefont {Hastings}, \citenamefont {Kothari},\ and\ \citenamefont
  {Low}}]{Haah}%
  \BibitemOpen
  \bibfield  {author} {\bibinfo {author} {\bibfnamefont {J.}~\bibnamefont
  {Haah}}, \bibinfo {author} {\bibfnamefont {M.~B.}\ \bibnamefont {Hastings}},
  \bibinfo {author} {\bibfnamefont {R.}~\bibnamefont {Kothari}}, \ and\
  \bibinfo {author} {\bibfnamefont {G.~H.}\ \bibnamefont {Low}},\ }\href
  {\doibase 10.1137/18M1231511} {\bibfield  {journal} {\bibinfo  {journal}
  {SIAM Journal on Computing}\ }\textbf {\bibinfo {volume} {0}},\ \bibinfo
  {pages} {FOCS18} (\bibinfo {year} {0})}\BibitemShut {NoStop}%
\bibitem [{\citenamefont {Childs}\ \emph {et~al.}(2021)\citenamefont {Childs},
  \citenamefont {Su}, \citenamefont {Tran}, \citenamefont {Wiebe},\ and\
  \citenamefont {Zhu}}]{Childs}%
  \BibitemOpen
  \bibfield  {author} {\bibinfo {author} {\bibfnamefont {A.~M.}\ \bibnamefont
  {Childs}}, \bibinfo {author} {\bibfnamefont {Y.}~\bibnamefont {Su}}, \bibinfo
  {author} {\bibfnamefont {M.~C.}\ \bibnamefont {Tran}}, \bibinfo {author}
  {\bibfnamefont {N.}~\bibnamefont {Wiebe}}, \ and\ \bibinfo {author}
  {\bibfnamefont {S.}~\bibnamefont {Zhu}},\ }\href {\doibase
  10.1103/PhysRevX.11.011020} {\bibfield  {journal} {\bibinfo  {journal} {Phys.
  Rev. X}\ }\textbf {\bibinfo {volume} {11}},\ \bibinfo {pages} {011020}
  (\bibinfo {year} {2021})}\BibitemShut {NoStop}%
\bibitem [{IBM(2021)}]{IBM}%
  \BibitemOpen
  \href {https://quantum-computing.ibm.com} {\enquote {\bibinfo {title} {{IBM
  Quantum} https://quantum-computing.ibm.com},}\ } (\bibinfo {year}
  {2021})\BibitemShut {NoStop}%
\bibitem [{\citenamefont {Foss-Feig}\ \emph {et~al.}(2020)\citenamefont
  {Foss-Feig}, \citenamefont {Hayes}, \citenamefont {Dreiling}, \citenamefont
  {Figgatt}, \citenamefont {Gaebler}, \citenamefont {Moses}, \citenamefont
  {Pino},\ and\ \citenamefont {Potter}}]{Potter}%
  \BibitemOpen
  \bibfield  {author} {\bibinfo {author} {\bibfnamefont {M.}~\bibnamefont
  {Foss-Feig}}, \bibinfo {author} {\bibfnamefont {D.}~\bibnamefont {Hayes}},
  \bibinfo {author} {\bibfnamefont {J.~M.}\ \bibnamefont {Dreiling}}, \bibinfo
  {author} {\bibfnamefont {C.}~\bibnamefont {Figgatt}}, \bibinfo {author}
  {\bibfnamefont {J.~P.}\ \bibnamefont {Gaebler}}, \bibinfo {author}
  {\bibfnamefont {S.~A.}\ \bibnamefont {Moses}}, \bibinfo {author}
  {\bibfnamefont {J.~M.}\ \bibnamefont {Pino}}, \ and\ \bibinfo {author}
  {\bibfnamefont {A.~C.}\ \bibnamefont {Potter}},\ }\href@noop {} {\enquote
  {\bibinfo {title} {Holographic quantum algorithms for simulating correlated
  spin systems},}\ } (\bibinfo {year} {2020}),\ \Eprint
  {http://arxiv.org/abs/2005.03023} {arXiv:2005.03023 [quant-ph]} \BibitemShut
  {NoStop}%
\end{thebibliography}%


\begin{thebibliography}{15}%
\makeatletter
\providecommand \@ifxundefined [1]{%
 \@ifx{#1\undefined}
}%
\providecommand \@ifnum [1]{%
 \ifnum #1\expandafter \@firstoftwo
 \else \expandafter \@secondoftwo
 \fi
}%
\providecommand \@ifx [1]{%
 \ifx #1\expandafter \@firstoftwo
 \else \expandafter \@secondoftwo
 \fi
}%
\providecommand \natexlab [1]{#1}%
\providecommand \enquote  [1]{``#1''}%
\providecommand \bibnamefont  [1]{#1}%
\providecommand \bibfnamefont [1]{#1}%
\providecommand \citenamefont [1]{#1}%
\providecommand \href@noop [0]{\@secondoftwo}%
\providecommand \href [0]{\begingroup \@sanitize@url \@href}%
\providecommand \@href[1]{\@@startlink{#1}\@@href}%
\providecommand \@@href[1]{\endgroup#1\@@endlink}%
\providecommand \@sanitize@url [0]{\catcode `\\12\catcode `\$12\catcode
  `\&12\catcode `\#12\catcode `\^12\catcode `\_12\catcode `\%12\relax}%
\providecommand \@@startlink[1]{}%
\providecommand \@@endlink[0]{}%
\providecommand \url  [0]{\begingroup\@sanitize@url \@url }%
\providecommand \@url [1]{\endgroup\@href {#1}{\urlprefix }}%
\providecommand \urlprefix  [0]{URL }%
\providecommand \Eprint [0]{\href }%
\providecommand \doibase [0]{http://dx.doi.org/}%
\providecommand \selectlanguage [0]{\@gobble}%
\providecommand \bibinfo  [0]{\@secondoftwo}%
\providecommand \bibfield  [0]{\@secondoftwo}%
\providecommand \translation [1]{[#1]}%
\providecommand \BibitemOpen [0]{}%
\providecommand \bibitemStop [0]{}%
\providecommand \bibitemNoStop [0]{.\EOS\space}%
\providecommand \EOS [0]{\spacefactor3000\relax}%
\providecommand \BibitemShut  [1]{\csname bibitem#1\endcsname}%
\let\auto@bib@innerbib\@empty
\bibitem [{\citenamefont {Moudgalya}\ \emph {et~al.}(2019)\citenamefont
  {Moudgalya}, \citenamefont {Prem}, \citenamefont {Nandkishore}, \citenamefont
  {Regnault},\ and\ \citenamefont {Bernevig}}]{MoudgalyaKrylov}%
  \BibitemOpen
  \bibfield  {author} {\bibinfo {author} {\bibfnamefont {S.}~\bibnamefont
  {Moudgalya}}, \bibinfo {author} {\bibfnamefont {A.}~\bibnamefont {Prem}},
  \bibinfo {author} {\bibfnamefont {R.}~\bibnamefont {Nandkishore}}, \bibinfo
  {author} {\bibfnamefont {N.}~\bibnamefont {Regnault}}, \ and\ \bibinfo
  {author} {\bibfnamefont {B.~A.}\ \bibnamefont {Bernevig}},\ }\href@noop {}
  {\enquote {\bibinfo {title} {Thermalization and its absence within krylov
  subspaces of a constrained hamiltonian},}\ } (\bibinfo {year} {2019}),\
  \Eprint {http://arxiv.org/abs/1910.14048} {arXiv:1910.14048
  [cond-mat.str-el]} \BibitemShut {NoStop}%
\bibitem [{\citenamefont {Yang}\ \emph {et~al.}(2012)\citenamefont {Yang},
  \citenamefont {Papi\ifmmode~\acute{c}\else \'{c}\fi{}}, \citenamefont
  {Rezayi}, \citenamefont {Bhatt},\ and\ \citenamefont
  {Haldane}}]{BoYangPhysRevB.85.165318}%
  \BibitemOpen
  \bibfield  {author} {\bibinfo {author} {\bibfnamefont {B.}~\bibnamefont
  {Yang}}, \bibinfo {author} {\bibfnamefont {Z.}~\bibnamefont
  {Papi\ifmmode~\acute{c}\else \'{c}\fi{}}}, \bibinfo {author} {\bibfnamefont
  {E.~H.}\ \bibnamefont {Rezayi}}, \bibinfo {author} {\bibfnamefont {R.~N.}\
  \bibnamefont {Bhatt}}, \ and\ \bibinfo {author} {\bibfnamefont {F.~D.~M.}\
  \bibnamefont {Haldane}},\ }\href {\doibase 10.1103/PhysRevB.85.165318}
  {\bibfield  {journal} {\bibinfo  {journal} {Phys. Rev. B}\ }\textbf {\bibinfo
  {volume} {85}},\ \bibinfo {pages} {165318} (\bibinfo {year}
  {2012})}\BibitemShut {NoStop}%
\bibitem [{\citenamefont {Papi\ifmmode~\acute{c}\else
  \'{c}\fi{}}(2013)}]{PapicTilt}%
  \BibitemOpen
  \bibfield  {author} {\bibinfo {author} {\bibfnamefont {Z.}~\bibnamefont
  {Papi\ifmmode~\acute{c}\else \'{c}\fi{}}},\ }\href {\doibase
  10.1103/PhysRevB.87.245315} {\bibfield  {journal} {\bibinfo  {journal} {Phys.
  Rev. B}\ }\textbf {\bibinfo {volume} {87}},\ \bibinfo {pages} {245315}
  (\bibinfo {year} {2013})}\BibitemShut {NoStop}%
\bibitem [{\citenamefont {Yang}\ \emph {et~al.}(2017)\citenamefont {Yang},
  \citenamefont {Lee}, \citenamefont {Zhang},\ and\ \citenamefont
  {Hu}}]{BoYangTilt}%
  \BibitemOpen
  \bibfield  {author} {\bibinfo {author} {\bibfnamefont {B.}~\bibnamefont
  {Yang}}, \bibinfo {author} {\bibfnamefont {C.~H.}\ \bibnamefont {Lee}},
  \bibinfo {author} {\bibfnamefont {C.}~\bibnamefont {Zhang}}, \ and\ \bibinfo
  {author} {\bibfnamefont {Z.-X.}\ \bibnamefont {Hu}},\ }\href {\doibase
  10.1103/PhysRevB.96.195140} {\bibfield  {journal} {\bibinfo  {journal} {Phys.
  Rev. B}\ }\textbf {\bibinfo {volume} {96}},\ \bibinfo {pages} {195140}
  (\bibinfo {year} {2017})}\BibitemShut {NoStop}%
\bibitem [{\citenamefont {Liu}\ \emph {et~al.}(2018)\citenamefont {Liu},
  \citenamefont {Gromov},\ and\ \citenamefont {Papi\ifmmode~\acute{c}\else
  \'{c}\fi{}}}]{PapicMain}%
  \BibitemOpen
  \bibfield  {author} {\bibinfo {author} {\bibfnamefont {Z.}~\bibnamefont
  {Liu}}, \bibinfo {author} {\bibfnamefont {A.}~\bibnamefont {Gromov}}, \ and\
  \bibinfo {author} {\bibfnamefont {Z.}~\bibnamefont
  {Papi\ifmmode~\acute{c}\else \'{c}\fi{}}},\ }\href {\doibase
  10.1103/PhysRevB.98.155140} {\bibfield  {journal} {\bibinfo  {journal} {Phys.
  Rev. B}\ }\textbf {\bibinfo {volume} {98}},\ \bibinfo {pages} {155140}
  (\bibinfo {year} {2018})}\BibitemShut {NoStop}%
\bibitem [{\citenamefont {Gromov}\ and\ \citenamefont {Son}(2017)}]{GromovSon}%
  \BibitemOpen
  \bibfield  {author} {\bibinfo {author} {\bibfnamefont {A.}~\bibnamefont
  {Gromov}}\ and\ \bibinfo {author} {\bibfnamefont {D.~T.}\ \bibnamefont
  {Son}},\ }\href {\doibase 10.1103/PhysRevX.7.041032} {\bibfield  {journal}
  {\bibinfo  {journal} {Phys. Rev. X}\ }\textbf {\bibinfo {volume} {7}},\
  \bibinfo {pages} {041032} (\bibinfo {year} {2017})}\BibitemShut {NoStop}%
\bibitem [{\citenamefont {Girvin}\ \emph {et~al.}(1985)\citenamefont {Girvin},
  \citenamefont {MacDonald},\ and\ \citenamefont {Platzman}}]{GMP85}%
  \BibitemOpen
  \bibfield  {author} {\bibinfo {author} {\bibfnamefont {S.~M.}\ \bibnamefont
  {Girvin}}, \bibinfo {author} {\bibfnamefont {A.~H.}\ \bibnamefont
  {MacDonald}}, \ and\ \bibinfo {author} {\bibfnamefont {P.~M.}\ \bibnamefont
  {Platzman}},\ }\href {\doibase 10.1103/PhysRevLett.54.581} {\bibfield
  {journal} {\bibinfo  {journal} {Phys. Rev. Lett.}\ }\textbf {\bibinfo
  {volume} {54}},\ \bibinfo {pages} {581} (\bibinfo {year} {1985})}\BibitemShut
  {NoStop}%
\bibitem [{\citenamefont {Carroll}\ and\ \citenamefont
  {Carroll}(2004)}]{carroll2004spacetime}%
  \BibitemOpen
  \bibfield  {author} {\bibinfo {author} {\bibfnamefont {S.}~\bibnamefont
  {Carroll}}\ and\ \bibinfo {author} {\bibfnamefont {S.}~\bibnamefont
  {Carroll}},\ }\href {https://books.google.co.uk/books?id=1SKFQgAACAAJ} {\emph
  {\bibinfo {title} {Spacetime and Geometry: An Introduction to General
  Relativity}}}\ (\bibinfo  {publisher} {Addison Wesley},\ \bibinfo {year}
  {2004})\BibitemShut {NoStop}%
\bibitem [{\citenamefont {Wen}\ and\ \citenamefont {Zee}(1992)}]{wen1992shift}%
  \BibitemOpen
  \bibfield  {author} {\bibinfo {author} {\bibfnamefont {X.~G.}\ \bibnamefont
  {Wen}}\ and\ \bibinfo {author} {\bibfnamefont {A.}~\bibnamefont {Zee}},\
  }\href {\doibase 10.1103/PhysRevLett.69.953} {\bibfield  {journal} {\bibinfo
  {journal} {Phys. Rev. Lett.}\ }\textbf {\bibinfo {volume} {69}},\ \bibinfo
  {pages} {953} (\bibinfo {year} {1992})}\BibitemShut {NoStop}%
\bibitem [{\citenamefont {Rahmani}\ \emph {et~al.}(2020)\citenamefont
  {Rahmani}, \citenamefont {Sung}, \citenamefont {Putterman}, \citenamefont
  {Roushan}, \citenamefont {Ghaemi},\ and\ \citenamefont {Jiang}}]{Rahmani}%
  \BibitemOpen
  \bibfield  {author} {\bibinfo {author} {\bibfnamefont {A.}~\bibnamefont
  {Rahmani}}, \bibinfo {author} {\bibfnamefont {K.~J.}\ \bibnamefont {Sung}},
  \bibinfo {author} {\bibfnamefont {H.}~\bibnamefont {Putterman}}, \bibinfo
  {author} {\bibfnamefont {P.}~\bibnamefont {Roushan}}, \bibinfo {author}
  {\bibfnamefont {P.}~\bibnamefont {Ghaemi}}, \ and\ \bibinfo {author}
  {\bibfnamefont {Z.}~\bibnamefont {Jiang}},\ }\href {\doibase
  10.1103/PRXQuantum.1.020309} {\bibfield  {journal} {\bibinfo  {journal} {PRX
  Quantum}\ }\textbf {\bibinfo {volume} {1}},\ \bibinfo {pages} {020309}
  (\bibinfo {year} {2020})}\BibitemShut {NoStop}%
\bibitem [{\citenamefont {Abraham}\ and\ \citenamefont
  {et~al.}(2019)}]{Qiskit}%
  \BibitemOpen
  \bibfield  {author} {\bibinfo {author} {\bibfnamefont {H.}~\bibnamefont
  {Abraham}}\ and\ \bibinfo {author} {\bibnamefont {et~al.}},\ }\href {\doibase
  10.5281/zenodo.2562110} {\enquote {\bibinfo {title} {Qiskit: An open-source
  framework for quantum computing},}\ } (\bibinfo {year} {2019})\BibitemShut
  {NoStop}%
\bibitem [{\citenamefont {Nam}\ \emph {et~al.}(2018)\citenamefont {Nam},
  \citenamefont {Ross}, \citenamefont {Su}, \citenamefont {Childs},\ and\
  \citenamefont {Maslov}}]{Optimiz_Q}%
  \BibitemOpen
  \bibfield  {author} {\bibinfo {author} {\bibfnamefont {Y.}~\bibnamefont
  {Nam}}, \bibinfo {author} {\bibfnamefont {N.~J.}\ \bibnamefont {Ross}},
  \bibinfo {author} {\bibfnamefont {Y.}~\bibnamefont {Su}}, \bibinfo {author}
  {\bibfnamefont {A.~M.}\ \bibnamefont {Childs}}, \ and\ \bibinfo {author}
  {\bibfnamefont {D.}~\bibnamefont {Maslov}},\ }\href@noop {} {\bibfield
  {journal} {\bibinfo  {journal} {npj Quantum Inf}\ }\textbf {\bibinfo {volume}
  {4}} (\bibinfo {year} {2018})}\BibitemShut {NoStop}%
\bibitem [{\citenamefont {Farhi}\ \emph {et~al.}(2014)\citenamefont {Farhi},
  \citenamefont {Goldstone},\ and\ \citenamefont {Gutmann}}]{farhi2014quantum}%
  \BibitemOpen
  \bibfield  {author} {\bibinfo {author} {\bibfnamefont {E.}~\bibnamefont
  {Farhi}}, \bibinfo {author} {\bibfnamefont {J.}~\bibnamefont {Goldstone}}, \
  and\ \bibinfo {author} {\bibfnamefont {S.}~\bibnamefont {Gutmann}},\
  }\href@noop {} {\enquote {\bibinfo {title} {A quantum approximate
  optimization algorithm},}\ } (\bibinfo {year} {2014}),\ \Eprint
  {http://arxiv.org/abs/1411.4028} {arXiv:1411.4028 [quant-ph]} \BibitemShut
  {NoStop}%
\bibitem [{\citenamefont {Saravanan}\ and\ \citenamefont
  {Saeed}(2022)}]{saravanan2022pauli}%
  \BibitemOpen
  \bibfield  {author} {\bibinfo {author} {\bibfnamefont {V.}~\bibnamefont
  {Saravanan}}\ and\ \bibinfo {author} {\bibfnamefont {S.~M.}\ \bibnamefont
  {Saeed}},\ }\href@noop {} {\enquote {\bibinfo {title} {Pauli error
  propagation-based gate reschedulingfor quantum circuit error mitigation},}\ }
  (\bibinfo {year} {2022}),\ \Eprint {http://arxiv.org/abs/2201.12946}
  {arXiv:2201.12946 [quant-ph]} \BibitemShut {NoStop}%
\bibitem [{\citenamefont {Davis}\ \emph {et~al.}(2020)\citenamefont {Davis},
  \citenamefont {Smith}, \citenamefont {Tudor}, \citenamefont {Sen},
  \citenamefont {Siddiqi},\ and\ \citenamefont {Iancu}}]{9259942}%
  \BibitemOpen
  \bibfield  {author} {\bibinfo {author} {\bibfnamefont {M.~G.}\ \bibnamefont
  {Davis}}, \bibinfo {author} {\bibfnamefont {E.}~\bibnamefont {Smith}},
  \bibinfo {author} {\bibfnamefont {A.}~\bibnamefont {Tudor}}, \bibinfo
  {author} {\bibfnamefont {K.}~\bibnamefont {Sen}}, \bibinfo {author}
  {\bibfnamefont {I.}~\bibnamefont {Siddiqi}}, \ and\ \bibinfo {author}
  {\bibfnamefont {C.}~\bibnamefont {Iancu}},\ }in\ \href {\doibase
  10.1109/QCE49297.2020.00036} {\emph {\bibinfo {booktitle} {2020 IEEE
  International Conference on Quantum Computing and Engineering (QCE)}}}\
  (\bibinfo {year} {2020})\ pp.\ \bibinfo {pages} {223--234}\BibitemShut
  {NoStop}%
\end{thebibliography}%
\onecolumngrid 
\newpage
\end{document}


\onecolumngrid 
\newpage

\begin{center}
{\bf \large Supplemental Material for ``Probing Geometric Excitations of Fractional Quantum Hall States on Quantum Computers"}
\end{center}
\begin{center}
Ammar Kirmani$^1$, Kieran Bull$^2$, Chang-Yu Hou$^3$, Vedika Saravanan$^4$, Samah Mohamed Saeed$^4$, Zlatko Papi\'c$^2$, Armin Rahmani$^{5,6}$, and Pouyan Ghaemi$^{7,8}$\\
\vspace*{0.1cm}
{\footnotesize
\sl 
$^1$Department of Physics and Astronomy, Western Washington University, Bellingham, Washington 98225, USA\\
$^2$School of Physics and Astronomy, University of Leeds, Leeds LS2 9JT, United Kingdom\\
$^3$Schlumberger-Doll Research, Cambridge, MA 02139, USA\\
$^4$Department of Electrical Engineering, City College of the City University of New York, NY 10031, USA\\
$^5$Department of Physics and Astronomy and Advanced Materials Science and Engineering Center, Western Washington University, Bellingham, Washington 98225, USA\\
$^6$Kavli Institute for Theoretical Physics, University of California, Santa Barbara, California 93106, USA\\
$^7$Physics Department, City College of the City University of New York, New York, NY 10031, USA\\
$^8$Graduate Center of the City University of New York, New York, NY 10016, USA

}
\end{center}

\setcounter{page}{1}
\setcounter{subsection}{0}
\setcounter{equation}{0}
\setcounter{figure}{0}
\renewcommand{\theequation}{S\arabic{equation}}
\renewcommand{\thefigure}{S\arabic{figure}}
\renewcommand{\thesubsection}{S\arabic{subsection}}

{\footnotesize In this Supplementary Material, we give a detailed derivation of the FQH Hamiltonian on a thin cylinder and in the presence of mass anisotropy. We briefly review the bimetric theory of FQH states and present additional evidence that the model studied in the main text captures the graviton dynamics. Finally, we discuss the Trotterization approach to simulating quantum dynamics in this model, and demonstrate its inadequacy in the currently available NISQ devices.
}

\vspace*{1cm}

\twocolumngrid

\section{FQH Hamiltonian and the thin-cylinder limit}\label{app:basic}

In this Section, we derive the parent Hamiltonian for the Laughlin $\nu{=}1/3$ state on a thin cylinder with general mass metric. We pick the Landau gauge, in which the single-electron orbitals are given by
\beg\label{landau_cyl}
\phi_j^c(\mathbf{r}) = \frac{1}{\sqrt{\pi^{1/2}L_2}}e^{i\kappa j y}e^{-\frac{(x- j\kappa )^2}{2}},
\en
where $j$ denotes the orbital index within the lowest Landau level (LLL) and $L_2$ is the circumference of the cylinder, with $\kappa=2\pi/L_2$ (we work in units $\ell_B{=}1$). The number of available single-electron orbitals is given by the magnetic flux as $N_\phi = (L_1 L_2)/(2\pi)$.

The field operator creating an electron at position $\mathbf{r}$ is given by
\begin{eqnarray}
\Psi^\dagger(\mathbf{r}) =\sum_j \phi_j^c (\mathbf{r}) c_j^\dagger,
\end{eqnarray}
where $c_j^\dagger$ is the creation operator which creates an electron at orbital $j$ within LLL (and similarly for the annihilation operator).  From the field operator, the density operator in momentum space is given by
\beg\label{rho_q}
\begin{split}
	\hat\rho(\bq) &=\int d^2r\; \Psi\dg(\br) \Psi(\br) e^{i\bq.\br} \\&=e^{-\frac{q_y^2+q_x^2}{4}}e^{\frac{i q_x q_y}{2}}\sum_j e^{i q_x \kappa j}c\dg_{j+\frac{q_y}{\kappa}}c_{j} \equiv F(\bq)\bar{\rho}(\bq),
\end{split}
\en
where, we introduced the LLL form factor $F(\bq)=e^{-\frac{q_y^2+q_x^2}{4}}$  and $\bar{\rho}(\bq)$ is the projected density operator.

Finally, the FQH Hamiltonian in momentum space takes the form
\beg\label{H_q}
\hat{H}=\frac{1}{N_\phi} \sum_{\bq}\bar\rho(-\bq)\bar{V}(\bq)\bar\rho(\bq),\quad \bar{V}(\bq)=[F(\bq)]^2 v(\bq),
\en
where $v(\bq)$ is the interaction potential. For the $\nu=1/3$ Laughlin state, the potential is given by $V_1$ Haldane pseudopotential and it takes the form $1-|\bq|^2$ in momentum space, which translates into the Trugman-Kivelson real space potential $\nabla ^2 \delta(\br)$. 

In the previous derivations, we have implictly assumed that the electron band mass tensor,
\beg
g=g_m=\begin{pmatrix}
	g_{11} & g_{12}\\
	g_{12} & g_{22}
\end{pmatrix}.
\en
is isotropic, i.e., $g_{11}=g_{22}=1$, $g_{12}=g_{21}=0$.  In a more general case, mass anisotropy modifies the single-electron wave functions in Eq.~(\ref{landau_cyl}), which in turn modifies the effective interaction matrix elements. In momentum space, the effect of band mass metric is given by: 
\begin{eqnarray}
\nonumber \bar{V}(g_m,\bq) &=& (1-g_{11} q_x^2-g_{22} q_y^2-2g_{12}q_x q_y)\\ &\times & \exp\left( -\frac{g_{11} q_x^2 +g_{22} q_y^2+2g_{12} q_x q_y}{2}\right).\quad\quad
\end{eqnarray}
To obtain the cylinder Hamiltonian for a general metric, we integrate out $q_x$ in  Eq.~(\ref{H_q}) and use the above expression for $\bar{V}(g_m,\bq)$. Dropping the overall multiplicative constant and making use of translation invariance, we obtain
\begin{eqnarray}
\hat{H}=&\sum_{l,m,j} V_{l,m} c\dg_{j+m}c_{j+m+l}c\dg_{j+l}c_{j}, \\
V_{l,m} &= \kappa^3\frac{m^2-(g_{11}g_{22}-g_{12}^2) l^2}{g_{11}^{3/2}} e^{-\kappa^2\frac{m^2+l^2-2ig_{12}lm}{2g_{11}}}, 
\end{eqnarray}
which can be simplified due to the unimodular property of the mass tensor,  $g_{11}g_{22}-g_{12}^2=1$,  due to the magnetic flux through the system being fixed (${\rm det}\; g =1$). We will use the following convenient reparametrization of $g$
\begin{eqnarray}
g = \left(
\begin{array}{cc}
 \cosh Q +\cos \phi  \sinh Q & \sin \phi  \sinh Q \\
 \sin \phi  \sinh Q & \cosh Q-\cos \phi  \sinh Q \\
\end{array}
\right),\;\;\;\;\;\label{eq:ghat0}
\end{eqnarray}
in terms of real numbers $Q$ and $\phi$, which represent the stretch and rotation in the plane. With these simplifications, the final Hamiltonian can be written as
\beg\label{hfullapp}
\begin{split}
\hat{H}=\sum_{j=0}^{N_\phi -1}& \sum_{k>|m|}V_{k,m}c_{j+m}\dg c_{j+k}\dg c_{j+m+k}c_j,\\ V_{k,m}\propto(k^2&-m^2)\exp\left(-\kappa^2\frac{(k^2+m^2-2ikmg_{12})}{2 g_{11}}\right).
\end{split}
\en

\begin{figure}[b]
	\centering
		\centering
		\includegraphics[width= \linewidth]{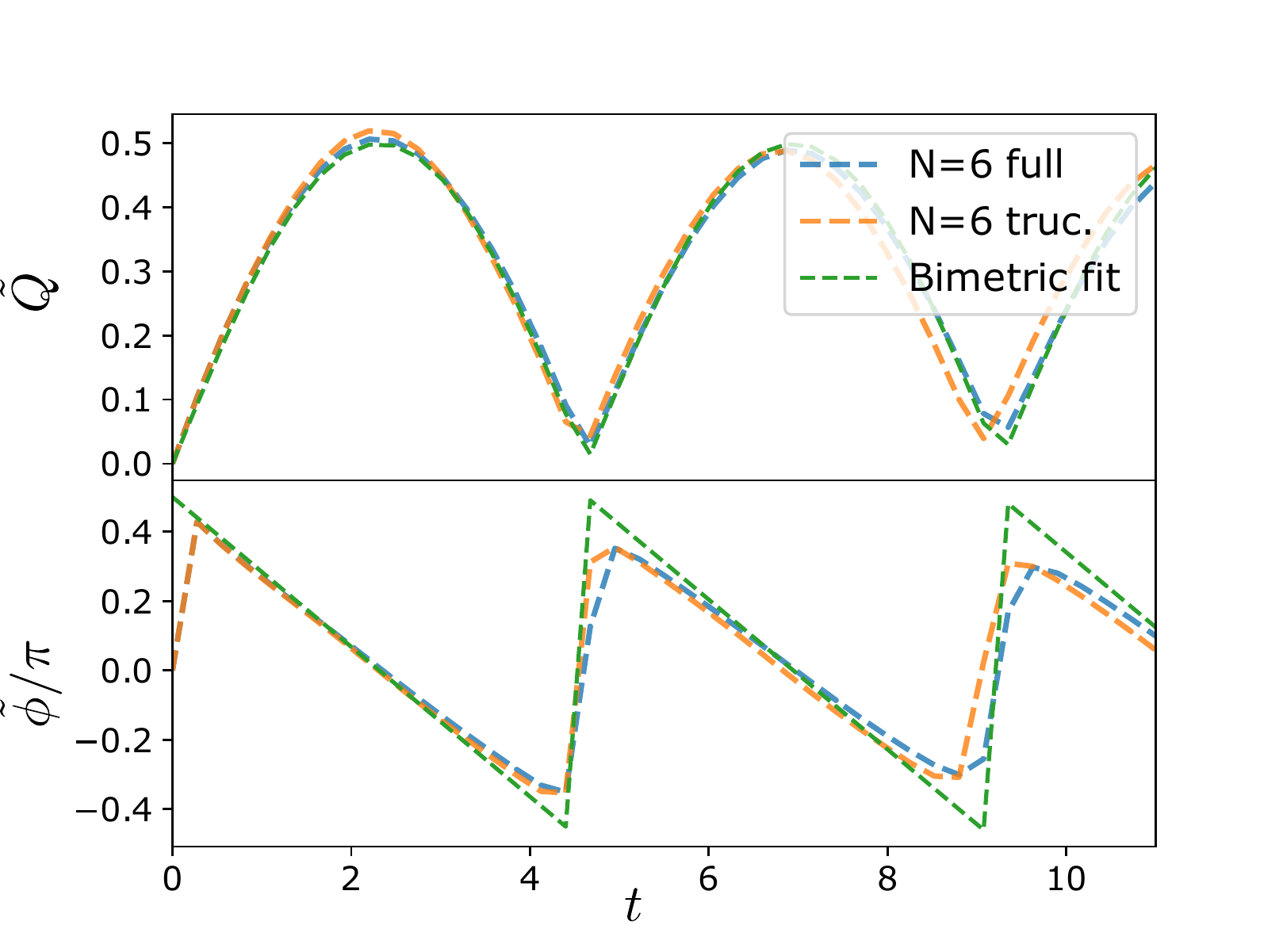}
		\caption{Comparison of the geometric quench dynamics between the full Hamiltonian, Eq.~(\ref{hfullapp}) and the truncated Hamiltonian, (\ref{Htruncapp}), for cylinder circumference $L_2{=}6.245$ and $N{=}6$ electrons. The quench is driven by changing $Q{=}0 \to Q{\approx}0.26$. Fit is against the bimetric theory prediction in Eq.~(\ref{eq:linsol1}) below. } \label{Fig:utrunc}
\end{figure}

In the thin torus limit, we can truncate the Hamiltonian in Eq.~(\ref{hfullapp}) by dropping long-range scattering processes, which are exponentially suppressed in the small parameter $\exp(-\kappa^2/2)$, arriving at the Hamiltonian studied in the main text,
\beg\label{Htruncapp}
\begin{split}
\hat{H} = \sum_{j} &
V_{1,0}\hat{n}_j \hat{n}_{j+1}+V_{2,0}\hat{n}_j \hat{n}_{j+2}+V_{3,0}\hat{n}_j \hat{n}_{j+3}
\\&+ V_{2,1}c\dg_{j+1}c\dg_{j+2}c_{j+3}c_{j} + \mathrm{h.c.}, 
\end{split}
\en
where $\hat{n}_j\equiv c_j\dg c_j$ and $V_{2,1}=3e^{-\kappa^2 \frac{5+i4 g_{12}}{2g_{11}}}$. Comparison of the dynamics generated by the Hamiltonian in Eq.~(\ref{hfullapp}) and the truncated Hamiltonian of Eq.~(\ref{Htruncapp}) are given in Fig \ref{Fig:utrunc}. We find that both models give approximately the same dynamics near the thin-cylinder limit, where we expect the truncation to be justified.

The Hamiltonian in Eq.~(\ref{Htruncapp}) is positive semi-definite for general $g_{12}$, so we can write it in the form  with 
\begin{eqnarray}\label{Model2}
\hat{H}&=&\sum_{j} ( {\cal Q}_j\dg {\cal Q}_j+{\cal P}_j\dg {\cal P}_j),\\
{\cal Q}_j &=& \sqrt{V_{1,0}} c_{j+2}c_{j+1}+\sqrt{V_{3,0}} e^{i\frac{8\pi^2 l_B^2}{L_2^2}\frac{g_{12}}{g_{11}}}c_j c_{j+3},\\ {\cal P}_j&=&\sqrt{V_{2,0}} c_j c_{j+2},
\end{eqnarray}
where $V_{1,0}$, $V_{2,0}$ and $V_{3,0}$ are given in Eq.~(\ref{hfullapp}) and $e^{i\frac{8\pi^2 l_B^2}{L_2^2}\frac{g_{12}}{g_{11}}}$ is the complex phase due to anisotropy parameterized by the previously mentioned tensor $g$.     
The ground state of Hamiltonian of Eq.~(\ref{Model2}) is given by the following expression given in the main text: 
\beg\label{SolSM}
\ket{\psi_{0}}={\cal N}\prod_{j} \left(1-\sqrt{\frac{V_{3,0}}{V_{1,0}}} e^{i\frac{8\pi^2 l_B^2}{L_2^2}\frac{g_{12}}{g_{11}}}\hat{S}_j \right)|...100100...\rangle,
\en
where $\hat{S}_j=c_{j+1}\dg c_{j+2}\dg c_{j+3} c_{j}$ is the 'squeezing' operator mentioned in the main text, and ${\cal N}$ is a normalization constant. It is evident that ${\cal Q}_j\ket{\psi_{0}}={\cal P}_j\ket{\psi_{0}}=0$ for all sites $j$,  and $\ket{\psi_{0}}$ is the ground state with zero energy. The geometric dependence enters our wave function in Eq.~(\ref{SolSM}) through the complex phase.

\section{Graviton excitation near the thin-cylinder limit}

The Hamiltonian in Eq.~(\ref{hfullapp}) has several symmetries, most importantly due to the specific form of $V_{k,m}$, it conserves the center-of-mass position  of the electrons, $K\equiv\sum_{j} j \hat{n}_j$. This is because each
$V_{k,m}$ term destroys two particles, initially separated by $|k-m|$ orbitals, and creates two particles at a distance $k+m$ orbitals (and vice versa). This means that, in the Landau gauge, the total momentum along the circumference is conserved. Thus, we can simultaneously label our energy states as momentum eigenstates denoted by quantum number $K$ (in the units of $2\pi/L_2$). It is convenient to label the orbitals $j=0,\pm 1, \ldots, \pm \frac{N_\phi-1}{2}$ such that $\sum j \hat{n}_j|...100100...\rangle=0$, i.e. the root state lies in the zero momentum sector. 

Before presenting the energy spectra of the models in Eqs.~(\ref{hfullapp})-(\ref{Htruncapp}), we explain the construction of their bases in Fock space. In the case of the model in Eq.~(\ref{hfullapp}), we work in the full Fock basis corresponding to $N$ spinless fermions residing in $N_\phi-2$ orbitals (note that the last two are always unoccupied sites). However, when considering the truncated Hamiltonian in Eq.~(\ref{Htruncapp}),  the effective Hilbert space is much smaller: it consists of  all configurations obtained by applying all possible squeezes to the root state. For example, the squeezed basis for $N{=}4$ electrons is $|100100100100\rangle$, $|011000100100\rangle$, $|100011000100\rangle$, $|100100011000\rangle$ and $|011000011000\rangle$. In general, the number of states in the squeezed basis is given by the Fibonacci number. Hence, the Hilbert space dimension still grows exponentially with system size, but it is much smaller than the full Fock basis. This is due to the special structure in the Hamiltonian~(\ref{Htruncapp}) which causes it to fracture into many dynamically-disconnected sectors~\cite{MoudgalyaKrylov}. 

\begin{figure}[htb]
\centering
		\centering
		\includegraphics[width= \linewidth]{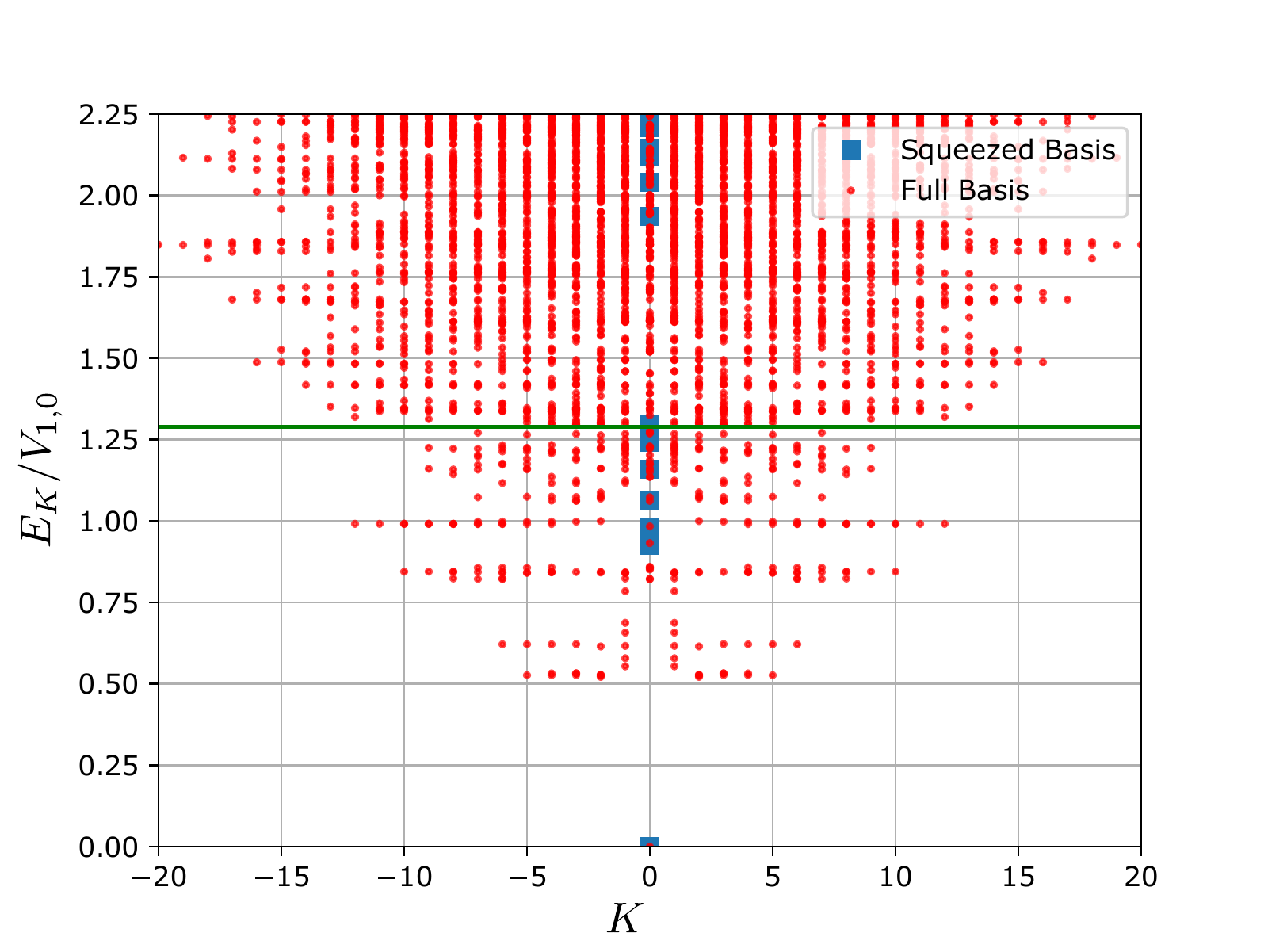}
		\caption{Spectrum of the Hamiltonian in Eq.~(\ref{Htruncapp}) with energies labelled by momentum $K$. Blue squares: spectrum obtained by restricting the Hamiltonian to the squeezed basis. Red dots: spectrum in the full Fock basis. The green line corresponds to the graviton energy.
		} \label{Fig:K}
\end{figure}
The energy spectrum $E_{K}$ of the quench Hamiltonian is plotted in Fig.~\ref{Fig:K}. We notice that the dominant state ($E_K{\approx}1.29$) contributing to the dynamics is much higher than the first excited state in $K{=}0$ momentum sector, for both the squeezed basis and the full Fock basis. Note that the excited states with momenta  $K{\neq}0$ do not contribute to the dynamics since our quench preserves translation symmetry (hence, $K$ is conserved). In the main text, we identified $E_g{\approx}1.29$ as the frequency of the emergent graviton mode which lies in the zero momentum sector. We have observed minor size and edge effects i.e. $E_g\approx1.26$ for $N=5$, $1.29$ for $N=7-9$ and $E_g\approx1.3$ for $N=15$.

\section{Breakdown of the bimetric theory in the strict 1D limit}

\begin{figure}[htb]
	\centering
		\centering
		\includegraphics[width= \linewidth]{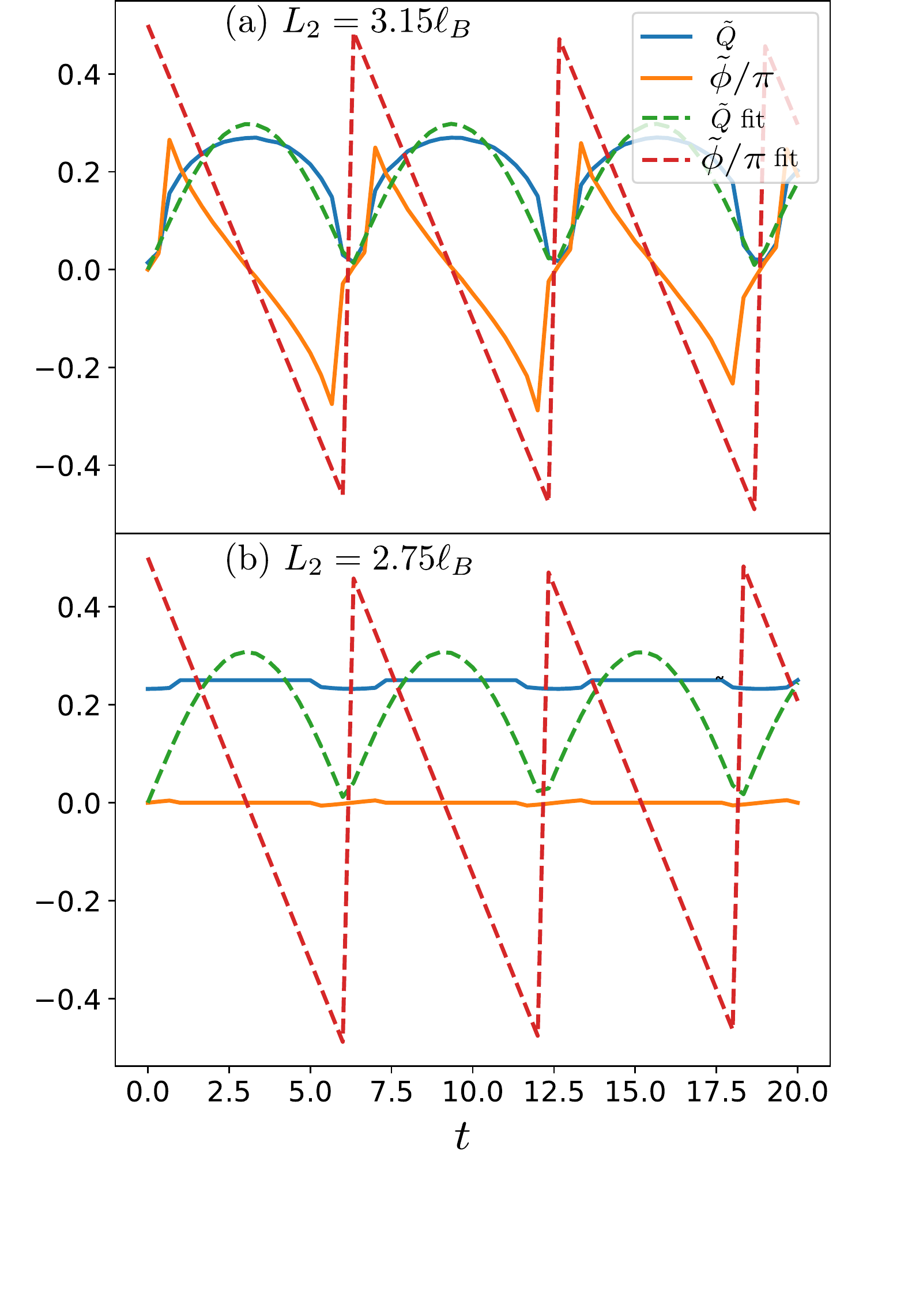}
		\caption{\color{black} Breakdown of graviton oscillations in the strict 1D limit. Plots show the deviation of microscopic dynamics of $Q(t)$ and $\phi(t)$ from the bimetric theory, for two values of the cylinder circumference. (a) The circumference is $3.15\ell_B$ and the dynamics still roughly follows the bimetic theory prediction, albeit with visible deviations. (b) The circumference is reduced to $2.75\ell_B$ and there is no agreement with the bimetric theory. For this value of the circumference, we are in the strict 1D Tao-Thouless limit where the dynamics is trivial.} \label{Fig:tt}
\end{figure}

For the observation of graviton oscillations in the main text, it was crucial that the FQH system is \emph{not} in the strict 1D limit, $L_2{\to}0$. We require a sufficiently large $L_2$, typically $L_2 \gtrsim 5\ell_B$, such that the system can accommodate the essential correlations in neutral excitations underpinning the graviton mode. Taking $L_2{\to}0$ destroys these correlations, the graviton oscillations disappear and the bimetric theory description breaks down, as we illustrate in Fig.~\ref{Fig:tt}. 

In Fig.~\ref{Fig:tt} we contrast the dynamics of $\tilde Q(t)$ and $\tilde \phi(t)$ for two values of the cylinder circumference, $L_2=3.15\ell_B$ (a) and $L_2=2.75\ell_B$ (b). While in the first case the dynamics still largely follows the bimetric theory prediction (albeit with visible deviations), in the second case the dynamics no longer conforms to the bimetric theory.  In the second case, we are in the regime of the 1D Tao-Thouless limit, where the dynamics is trivial due to the initial state being close a product state and an eigenstate of the post-quench Hamiltonian. Thus, we conclude that finite $L_2$ is necessary to observe the graviton oscillation.



\section{Geometric quench in the bimetric theory}\label{appnd:bi}

In FQH systems, ``geometry'' appears in three distinct guises. First, the form factor $F_m(\mathbf{q})=\exp(-g_m^{ab}q_aq_b\ell_B^2/4)$ is a function of the band mass tensor $g_m$, where 
we use the Einstein summation convention. As discussed in Eq.~(\ref{eq:ghat0}), the symmetric, unimodular tensor $g_m$ can be conveniently parametrized by $Q$ and $\phi$. Second, the interaction potential $V_\mathbf{q}$ in general depends on another rank-2 tensor $g_i$, characterizing the underlying solid state material ($g_i$ originates from the dielectric tensor of the material for the Coulomb interaction). Both $g_m$ and $g_i$ are set by \emph{extrinstic} experimental conditions. For simplicity, in this paper we study the model Hamiltonian for the Laughlin state, where we take $g_m=g_i$.

In the presence of extrinsic tensors $g_m$ and $g_i$, a FQH state develops its own, \emph{intrinsic} geometric degree of freedom. This intrinsic degree of freedom defines the shape of particle-flux composite droplets in the FQH state and can be thought of as a metric $\widetilde{g}$, which is similarly parametrized by Eq.~(\ref{eq:ghat0}). Intrinsic metric $\widetilde{g}$ is a ``compromise" between $g_m$ and $g_i$~\cite{BoYangPhysRevB.85.165318}, because $g_m$ and $g_i$ are physically independent and in general different from each other. The dynamics of $\widetilde{g}$ can be induced by tilting the magnetic field, which can be exactly modelled by $2\times 2$ anisotropic mass tensor for parabolic confining potential~\cite{BoYangPhysRevB.85.165318, PapicTilt, BoYangTilt}. This forms the key component of the geometric quench protocol~\cite{PapicMain}. 

Bimetric theory~\cite{GromovSon} describes gapped dynamics of a single spin-$2$ degree of freedom which is present in the collective excitation spectrum of all gapped FQH states. In the long-wavelength limit, this degree of freedom has a variational description in terms of the Girvin-MacDonald-Platzman (GMP) neutral excitation~\cite{GMP85}. The bimetric theory assumes the spin-2 excitation mode couples to an external electro-magnetic field and ambient geometry. Thus, the dynamical degree of freedom is the vielbein $\hat e^\alpha_i$ \cite{carroll2004spacetime} that ``squares'' to give the  intrinsic metric introduced above, $\widetilde g_{ij} = \hat e^\alpha_i\hat e^\beta_j \delta_{\alpha \beta}$. The inverse metric is given by $\hat G^{ij} = \hat E_\alpha^i\hat E_\beta^j \delta^{\alpha \beta}$, where $\hat E_\alpha^i$ is the inverse vielbein, satisfying $\hat E^i_\alpha \hat e_j^\alpha = \delta^i_j$. The unimodular condition on $\widetilde g_{ij}$ reads $\sqrt{g} = \sqrt{\widetilde g}$, where $\sqrt{g}$ is the determinant of the ambient metric $g_{ij}$. Quadrupolar anisotropy is introduced by modifying the ambient metric,  $
g^{\mathrm{    m}}_{ij} = \mathrm{m}_{AB} e^A_i e^B_j
$, where $e^A_i$ are the vielbeins that describe the ambient geometry (we assume ambient space is flat, i.e., $e^A_i = \delta^A_i$) and 
$\mathrm{m}_{AB}$ is a unimodular matrix, assumed to be of the form ${\rm diag}\left[e^A, e^{-A}\right]$, where $A$ is the effective anisotropy.

In a homogeneous magnetic field, the bimetric Lagrangian is~\cite{GromovSon}
\begin{eqnarray}
\label{eq:isoL}
\mathcal L  = \frac{\nu \varsigma}{2\pi \ell_B^2}  \hat \omega_0 - \frac{m}{2} \left( \frac{1}{2}\widetilde g^{ij} g^\mathrm{m}_{ij} - \gamma \right)^2 \,,
\end{eqnarray}  
where $\hat \omega_0 = \frac{1}{2} \epsilon_{\alpha}{}^\beta \hat E^i_\beta \partial_0  \hat e_i^\alpha$ is the temporal component of the (dynamic) Levi-Civita spin connection.
The phenomenological coefficient $m$ sets the energy scale which determines the gap of the spin-$2$ mode,  $E_g = 2\Omega(1-\gamma)$, where $\Omega = (m/\varsigma)(2\pi\ell_B^2/\nu)$, and the quantized coefficient $\varsigma$ is determined by the ``shift"~\cite{wen1992shift}. Parameter $\gamma$ is used to tune the theory close to the nematic phase transition in the gapped phase $\gamma < 1$, where the GMP description is exact. 

In order to compute the dynamics of $\widetilde g$, we parametrize it in terms of $\widetilde{Q}$ and $\widetilde \phi$ as in Eq.~(\ref{eq:ghat0}).
Both $\widetilde Q$ and $\widetilde \phi$ are functions of time but not space, since we consider global (homogeneous) quench. The equations for $\widetilde \phi (t)$ and $\widetilde Q(t)$ have been shown to take the form~\cite{PapicMain}
\begin{eqnarray}
 \label{eq:phi}
\nonumber \dot{\widetilde \phi} \sinh \widetilde Q &=& - 2 \Omega \left( \sinh A \cosh \widetilde Q \cos \widetilde \phi -\cosh A \sinh \widetilde Q\right) \\
  &\times& \left(\gamma+\sinh A \sinh \widetilde Q \cos \phi -\cosh A \cosh \widetilde Q \right),\quad\quad  \\
\label{eq:Q}
\nonumber \dot{\widetilde Q} \sinh \widetilde Q &=& - 2 \Omega \sin \widetilde \phi \sinh \widetilde Q \sinh A  \\
&& \left( \gamma+\sinh A \sinh \widetilde Q \cos \widetilde \phi -\cosh A \cosh \widetilde Q \right).\quad \quad
\end{eqnarray}
This non-linear classical system governs the  universal dynamics of FQH states following a geometric quench. When anisotropy is weak, we can assume both $A$ and $\widetilde Q$ are close to $0$. By Taylor expanding Eqs.~(\ref{eq:phi}) and (\ref{eq:Q}) in $A$ and $\widetilde Q$, it can be shown that 
for a particular initial condition $\widetilde Q(0)=0$, the analytical solution is given 
\begin{eqnarray}
\widetilde Q(t) &=& \pm 2A\sin\left(\frac{E_g t}{2}\right),\;\;\; \widetilde \phi(t) = \pi \mp \frac{\pi}{2} -\frac{E_g t}{2},\;\;\;\;\label{eq:linsol1}
\end{eqnarray} 
which was used for fitting the dynamics data in the main text.  Note that there is only one linearly independent solution because the system is invariant under $\widetilde Q\rightarrow -\widetilde Q$ and $\widetilde \phi\rightarrow \widetilde \phi+\pi$. Thus, we can focus on the $\widetilde Q\geq0$ part and consider $\widetilde \phi\;{\rm mod}\;2\pi$. By inspection of $\widetilde Q(t)$, we see the solution alternates between two branches, which doubles the frequency from $E_g/2$ to $E_g$. The overall prefactor (written as $2A$) is expected to be proportional to the anisotropy of the post-quench Hamiltonian.

Ref.~\cite{PapicMain} established a good quantitative agreement between the numerically exact quench dynamics and the above prediction of the bimetric theory. The agreement is observed in the regime where the ground state before and after the quench remains in the Laughlin phase. Intuitively, the agreement is due to the fact that the quench involves an exponentially vanishing fraction of eigenstates that contribute to the dynamics, which \emph{a posteriori} justifies the fundamental assumption of the bimetric theory that assumes a single spin-$2$ degree of freedom.

\section{\bf Derivation of the spin-chain Hamiltonian} The Hamiltonian in Eq.\eqref{Htruncapp} can be mapped onto the spin model in Eq. \eqref{fullspapp} in terms of the reduced registers. Each reduced register corresponds to a block of three consecutive sites, where the reduced register $\ell$ contains three fermionic sites $j=3\ell-3,3\ell-2,3\ell-1$ as shown below: 
\begin{center}\includegraphics[width=4cm]{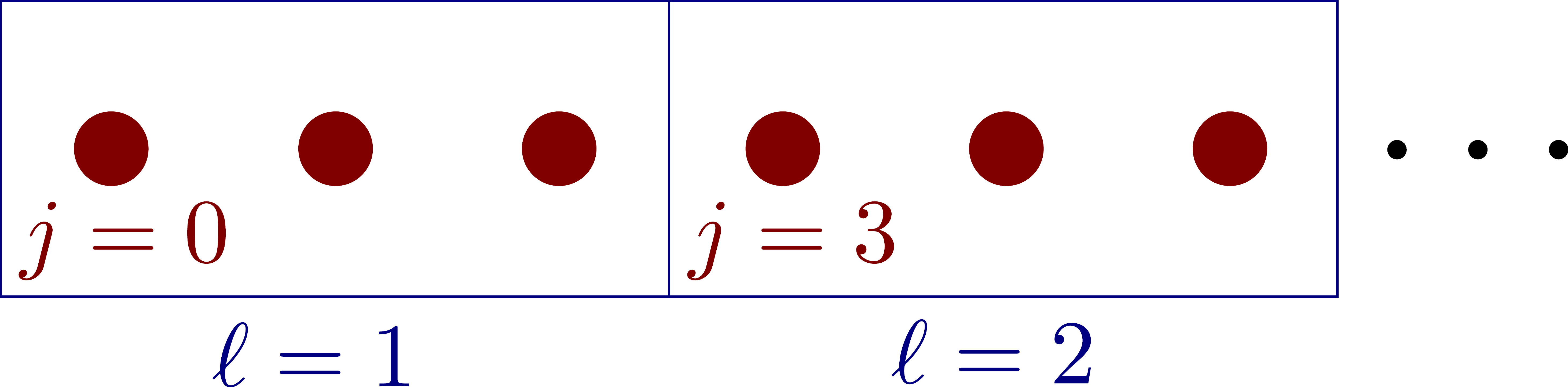} 
\end{center}
To derive the mapping, note that it is sufficient to restrict to the subspace of states obtained by repeated action of the Hamiltonian on the root state, i.e., $H |\ldots 100100\ldots\rangle, H^2 |\ldots 100100\ldots\rangle, $ etc., or, equivalently, states obtained by repeated action of the squeezing operators $\hat{S}_j=c\dg_{j+1}c\dg_{j+2}c_{j+3}^{}c_j^{}$ on the root state. We refer to this subspace as the squeezing subspace. Within this subspace, the terms $\hat{n}_j \hat{n}_{j+2}$ can be dropped as none of the states generated by the application of $\hat{S}_j$ on  $|...100100...\rangle$ have second neighbor 1s, hence $\hat{n}_j \hat{n}_{j+2}$ trivially vanishes in this subspace. 

Next, we consider the $\hat{n}_j \hat{n}_{j+1}$ terms. We focus on open boundary conditions, where the last block is never squeezed, i.e., ${\cal N}_N=\mathbb{0}$. Therefore $\ell{=}N{-}1$ is effectively the last register of the spin chain and register $\ell{=}N$ can be treated a ghost site. It is easy to observe that each squeezed block creates a pair of occupied nearest neighbors. For example, the state
$|\mathbb{10100}\rangle\sim|011, 000, 011, 000, 100\rangle$ 
has exactly two $\mathbb 1$ registers and two nearest neighbor  occupied pairs. Therefore,
\[
\sum_{j=0}^{3N-2}\hat{n}_j \hat{n}_{j+1} \to\sum_{\ell=1}^{N-1}{\cal N}_\ell.
\]

We next consider the $\hat{n}_j \hat{n}_{j+3}$ terms. The root state $|\mathbb{00000}\rangle=|100, 100, 100, 100, 100\rangle$, here shown for $N{=}5$, has $\sum_j\hat{n}_j \hat{n}_{j+3} =N{-}1$. Now let us consider adding a squeezed block $\ell$, somewhere in the middle of the chain, surrounded by unsqueezed blocks:
\begin{eqnarray}
...\mathbb{0000}...&=&...100, 100, 100, 100, 100...\nonumber \\
...\mathbb{0100}...&=&...100, 011, 000, 100, 100...
\end{eqnarray}
This operation changes blocks $\ell$ and $\ell+1$, reducing $\hat{n}_j \hat{n}_{j+3}$ by 3 as the bonds between blocks $(\ell-1, \ell)$, $(\ell, \ell+1)$, and $(\ell+1, \ell+2)$ no longer contribute to $\sum_j\hat{n}_j \hat{n}_{j+3}$. Therefore, we must subtract $3 \sum_\ell {\cal N}_\ell$ from the sum. However, if we squeeze two second-neighbor blocks $\ell$ and $\ell+1$ as
\[
...\mathbb{010100}...=..100, 011, 000, 011, 000, 100...
\]
instead of reducing $\hat{n}_j \hat{n}_{j+3}$ by $2\times 3=6$, we reduce it by 5 as the $(\ell+1, \ell+2)$ bond is double counted. We thus need to add a term $\sum_\ell {\cal N}_\ell{\cal N}_{\ell+2}$. The only other correction involves $\ell{=}1$ and $\ell{=}N{-}1$, which respectively do not have the $(\ell-1, \ell)$, and $(\ell+1, \ell+2)$ blocks and therefore only reduce the sum by 2 instead of 3. We then obtain
\[
\sum_{j=0}^{3N-4}\hat{n}_j \hat{n}_{j+3} \to N-1-3\sum_{\ell=1}^{N-1}{\cal N}_\ell+\sum_{\ell=1}^{N-3}{\cal N}_\ell{\cal N}_{\ell+2}+{\cal N}_1+{\cal N}_{N-1}.
\]

Finally, we map the squeezing (off-diagonal) term. Unless $j$ is the first site of a reduced block, the operator $V_{2,1} c\dg_{j+1}c\dg_{j+2}c_{j+3}c_j+ V_{2,1}^* c\dg_jc\dg_{j+3}c_{j+2}c_{j+1}$ annihilates the state. Thus we focus on an individual term corresponding to block $\ell$
\[
W_{\ell}=V_{2,1} c\dg_{j+1}c\dg_{j+2}c_{j+3}c_j+V_{2,1}^* c\dg_jc\dg_{j+3}c_{j+2}c_{j+1},
\]
where $j=3\ell-3$ is the first site of the block $\ell$. We notice that if either of the neighboring blocks of an unsqueezed block $\ell$ are squeezed, $W_\ell$ annihilates a state:
\begin{eqnarray}
W_\ell|...\mathbb{1}\mathbb{0}_\ell\mathbb{0}...\rangle&=&W_\ell|...011,000,100..\rangle=0,\nonumber \\
W_\ell|...\mathbb{0}\mathbb{0}_\ell\mathbb{1}...\rangle&=&W_\ell|...100,100,011..\rangle=0,\nonumber \\
W_\ell|...\mathbb{1}\mathbb{0}_\ell\mathbb{1}...\rangle&=&W_\ell|...011,000,011..\rangle=0.
\end{eqnarray} 
We can implement this annihilation by including a factor $(1-{\cal N}_{\ell-1})(1-{\cal N}_{\ell+1})$. Blocks $\ell=1$ and $\ell=N-1$ are special because they have only one neighboring block.
If block $\ell$ is already squeezed it cannot have a squeezed neighbor, in which case
\begin{eqnarray}
W_\ell|...\mathbb{0}\mathbb{0}_\ell\mathbb{0}...\rangle&=&V_{2,1} |...100,011,000..\rangle=V_{2,1} |...\mathbb{0}\mathbb{1}_\ell\mathbb{0}...\rangle,\nonumber \\
W_\ell|...\mathbb{0}\mathbb{1}_\ell\mathbb{0}...\rangle&=&V_{2,1}^* |...100,100,100..\rangle=V_{2,1}^* |...\mathbb{0}\mathbb{0}_\ell\mathbb{0}...\rangle.\nonumber
\end{eqnarray}
Therefore if the neighbors are $\mathbb 0$ the action of $W_\ell$ in the $(|\mathbb{0}\rangle, |\mathbb{1}\rangle)$ basis is given by the matrix
\[
W_\ell = \left(\begin{array}{cc}
0 & V_{2,1} \\ 
V_{2,1}^* & 0
\end{array} \right)={\rm Re}(V_{2,1})X-{\rm Im}(V_{2,1}) Y.
\]
We then represent the full squeezing term for open boundary condition as
\beg\label{fullspapp}
\begin{split}
&\sum_{j=0}^{3N-4} V_{2,1} c\dg_{j+1}c\dg_{j+2}c_{j+3}c_j+{V_{2,1}}^* c\dg_jc\dg_{j+3}c_{j+2}c_{j+1}\\&\to [{\rm Re}(V_{2,1})X_1-{\rm Im}(V_{2,1}) Y_1](1-{\cal N}_2)\\&+(1-{\cal N}_{N-2})[{\rm Re}(V_{2,1})X_{N-1}-{\rm Im}(V_{2,1}) Y_{N-1}]\\
&+\sum_{\ell=2}^{N-2}(1-{\cal N}_{\ell-1})[{\rm Re}(V_{2,1})X_\ell-{\rm Im}(V_{2,1}) Y_\ell](1-{\cal N}_{\ell+1}).
\end{split}
\en

\section{Details of Trotterization circuit}\label{appnd:trot}

{\color{black}
 In the main text, we wrote the quench Hamiltonian as 
 \[
 \hat H=\sum_\ell H_\ell,
 \]
 giving rise to a unitary operator
$
 U(t)=\left[\prod_\ell U_\ell (\delta t)\right]^k.
$
Here we explicitly write the $U_\ell$ operators, ignoring the commutations of the terms in the exponent in the limit of small $\delta t$. For $\ell=0$, we can write
\begin{align*}
    U_0 (\delta t)  \approx e^{-i \frac{V_{1,0} - 2 V_{3,0}}{2}(I + Z_0)\delta t} e^{-i V_{2,1}  X_0  (1-\mathcal{N}_1)\delta t},
\end{align*}
where $I$ is the identity. Ignoring a global phase, the first terms is a $z$ rotation and the second term is a controlled $x$ rotation, resulting in the circuit element of Fig. \ref{fig:U0}.
\begin{figure}[h]
    \centering
    \includegraphics[width=0.7\linewidth]{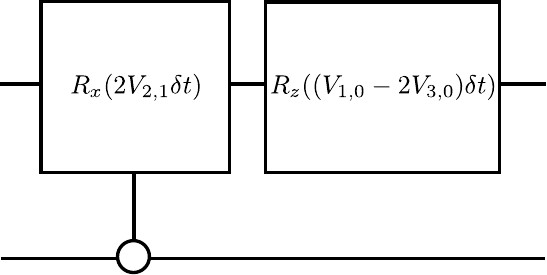}
    \caption{Circuit implementation of the Trotterized unitary $U_0$, at the boundary at the left of the chain.}
    \label{fig:U0}
\end{figure}
Similarly the unitary at the right boundary can be written as
\begin{align*}
    U_{N-1} (\delta t)  \approx e^{-i \frac{V_{1,0} - 2 V_{3,0}}{2}(I + Z_{N-1})\delta t} e^{-i V_{2,1} (1-\mathcal{N}_{N-2}) {X}_{N-1} \delta t},
\end{align*}
giving the circuit element shown in Fig. \ref{fig:U_N-1}.
\begin{figure}[h]
    \centering
    \includegraphics[width=0.7\linewidth]{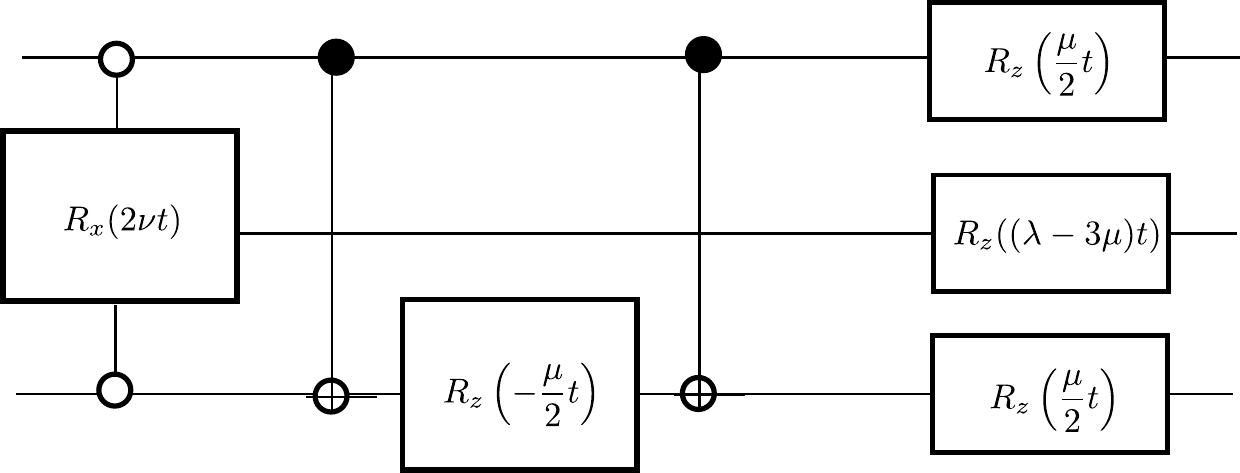}
    \caption{Circuit implementation of the Trotterized unitary $U_{N-1}$, at the boundary at the right of the chain.}
    \label{fig:U_N-1}
\end{figure}
 
In the bulk of the spin chain, the unitary operator up a global phase has the following structure 
\begin{align*}
    U_\ell(\delta t) &\approx  e^{-i \frac{V_{3,0}}{4} Z_{\ell-1}\delta  t} e^{-i \frac{V_{1,0} - 3 V_{3,0}}{2} Z_\ell \delta t} e^{-i \frac{V_{3,0}}{4} Z_{\ell+1}\delta t} \\
        &\times e^{-i \frac{V_{3,0}}{4} Z_{\ell-1} Z_{\ell+1} \delta t} e^{-i V_{2,1} (1-\mathcal{N}_{\ell l-1}) X_\ell (1-\mathcal{N}_{\ell+1})\delta t}. 
\end{align*}
Noting that the operator $e^{-i \frac{V_{3,0}}{4} Z_{\ell-1} Z_{\ell+1}\delta t}$ can be implemented by a $z$ rotation acting on qubit $\ell+1$ sandwiched between two {\small CNOT} gates with qubit $\ell-1$ as the control bit, we find the circuit element presented in the main text for implementing $U_\ell(\delta t)$ in the bulk. 
The unitary constructed above should then act on the prequench ground state. We use the circuit of Ref.~\onlinecite{Rahmani} to create the Laughlin $\nu=1/3$ fractional quantum Hall state as the initial state.}

\section{ Variational quantum algorithm.}
In the geometric quench, the initial state is the ground state $|\psi_0\rangle$ with the identity metric ($g_{12}{=}0$). Ref.~\cite{Rahmani} found an efficient algorithm to generate $|\psi_0\rangle$ with a linear-depth circuit. Working in the reduced space, we can use only stage 1 of the circuit in Ref.~\onlinecite{Rahmani} identifying qubits $1{+}3n$ with the reduced registers.  The state after the quench is generated by the unitary evolution operator $U(t){=}\exp(-it \hat H)$, where $\hat H$ is the anisotropic post-quench Hamiltonian. Thus, combining an algorithm that generates $U(t)$ with the the ground-state preparation algorithm yields the post-quench state $|\psi(t)\rangle{=}U(t)|\psi_0\rangle$.

\begin{figure}
	\centering
		\centering
		\includegraphics[width= \linewidth]{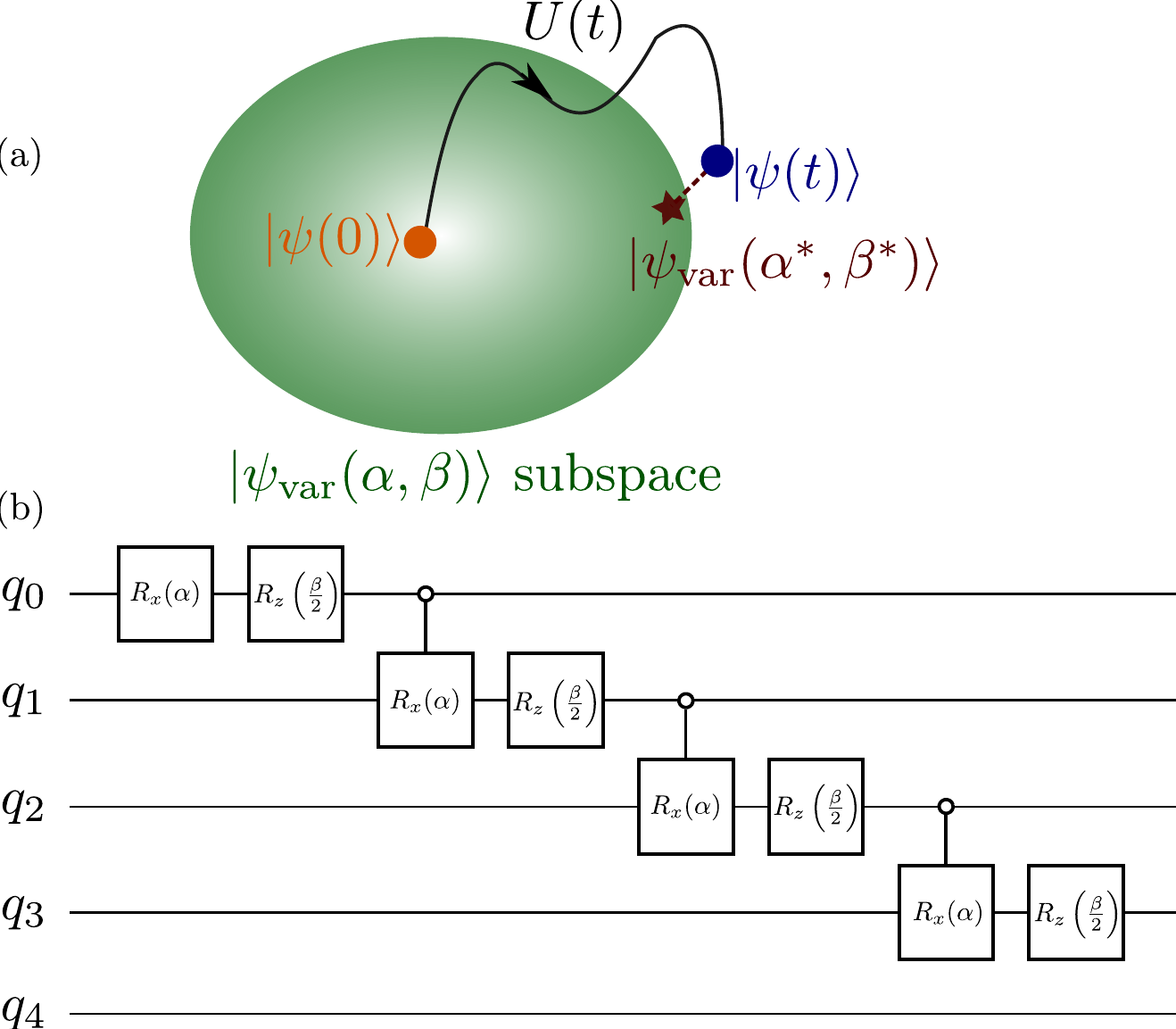}
		\caption{(a) The post-quench dynamics can be accurately approximated by a two-parameter variational ansatz $|\psi_{\rm var}(\alpha, \beta)\rangle$, which spans a small region of the Hilbert space. For every time $t$, there are optimal variational parameters $\alpha^*$ and $\beta^*$ for which the nonequilibrium wave function has an extremely high overlap with the variational ansatz. The optimal parameters have a smooth dependence on time and weak dependence on system size, providing excellent extrapolation to the thermodynamic limit. (b) The variational ansatz for the optimal parameters can be implemented in an efficient algorithm of linear circuit depth.}  \label{fig:var_schem}
\end{figure}
\begin{figure}[t]
	\centering
		\centering
		\includegraphics[width= \linewidth]{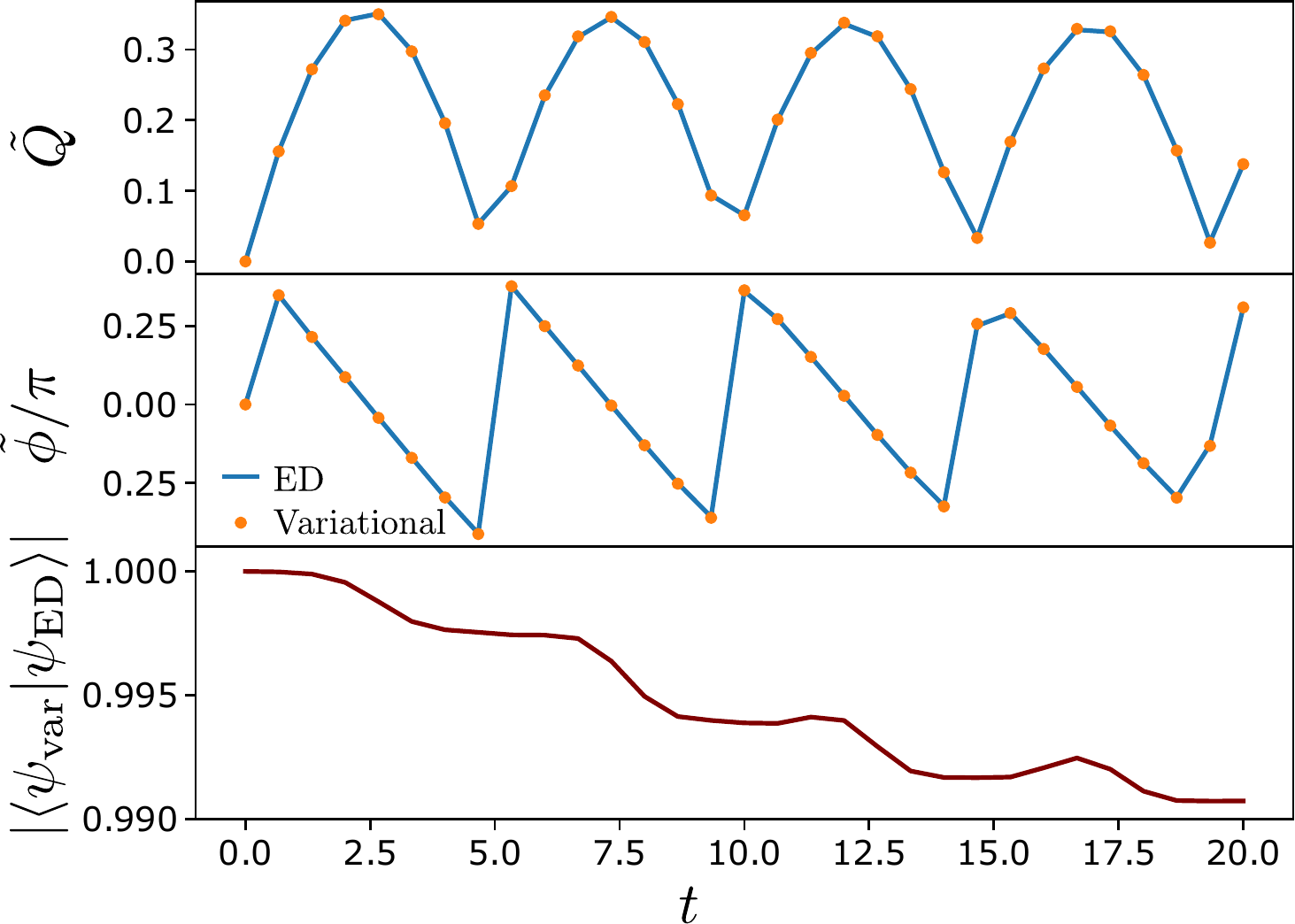}
		\caption{
		Variational ansatz in Eq.~(\ref{eq:ansatz}) successfully captures the time-dependent quantum metric parameters, $\widetilde Q$ and $\widetilde \phi$, as well as the full wave function obtained from exact diagonalization (ED) for $N{=}9$ electrons.} \label{Fig:VarEd}
\end{figure}

A large number of Trotterization steps is needed to obtain accurate results for $U(t)$. Furthermore, terms like $(1-{\cal N}_{\ell-1})X_\ell(1-{\cal N}_{\ell+1})$, which appear in  Eq.~(\ref{fullspapp}), require $R_x$ gates controlled by two other qubits, in turn requiring six {\small CNOT} gates for each control gate. In order to overcome these difficulties and scale the algorithm to larger systems, we approximate $|\psi(t)\rangle$ using the ansatz:
\begin{eqnarray}\label{eq:ansatzbig}
 |\psi_{\rm var}(\alpha_{i},\beta_{i})\rangle= \prod_\ell e^{-i\alpha_l {\cal N}_\ell} e^{-i\beta_l (1-{\cal N}_{\ell-1})X_\ell}|\mathbb{000}\dots\rangle, \quad 
\end{eqnarray}
The above anstaz has site-dependent variational parameters $\alpha_\ell, \beta_\ell$. As we have translational symmetry away from the boundaries, we can further simply this ansatz ans approximate the dynamics by only two variational parameters:
\begin{eqnarray}\label{eq:ansatz}
 |\psi_{\rm var}(\alpha,\beta)\rangle= \prod_\ell e^{-i\alpha {\cal N}_\ell} e^{-i\beta (1-{\cal N}_{\ell-1})X_\ell}|\mathbb{000}\dots\rangle, \quad
\end{eqnarray}
Indeed, we have numerically found that imposing  translation invariance on the variational parameters produces high overlaps (${\sim}0.99$) with the exact state and is sufficient for capturing any local observable expectation values in the bulk of the system, in particular it faithfully reproduces the quantum metric parameters $\widetilde Q$ and $\widetilde \phi$, Fig.~\ref{Fig:VarEd}.


The ansatz in Eq.~(\ref{eq:ansatz}) has several key advantages that reduce its implementation cost on NISQ hardware, while at the same time it captures any local observable expectation values in the bulk of the system, in particular it faithfully reproduces the quantum metric dynamics, $\widetilde Q(t)$ and $\widetilde \phi(t)$.
Importantly, the optimal parameters $\alpha^*$, $\beta^*$ are found to exhibit a simple oscillatory behavior as a function of time as mentioned in the main text. 

One advantage of the ansatz in Eq.~(\ref{eq:ansatz}) is that it relies only on two-qubit control gates, while ensuring no forbidden states (with neighboring $\mathbb{11}$ registers) are generated as shown in the Fig. \ref{fig:var_schem}.   Another advantage is that we can obtain the final state by acting on the trivial root state $|\mathbb{000}\dots\rangle$ rather than  the ground state $|\psi_0\rangle$. Since all registers are zero in the root state, if we start from the left end of the chain and apply controlled rotations, the qubit to the right is always zero before the application of the unitary operator, allowing us to use two-qubit control gates instead of three-qubit ones, with each register controlled only by its left neighbor. 


\section{Extrapolation of variational parameters}\label{appnd:extrap}


The weak system-size dependence of the optimal variational parameters $\alpha^*$ and $\beta^*$ allows us to extrapolate them to the thermodynamic limit, see Fig. 5 of the main text. We use $1/N$ as a small parameter, and assume $\alpha^*$ and $\beta^*$ have a Taylor expansion in powers of $1/N$ for each total time $t$:
\begin{eqnarray}
\alpha^*(N,t)&\approx &p_0^\alpha(t)+ \frac{p_1^\alpha(t)}{N}+\frac{p_2^\alpha(t)}{N^2}+\frac{p_3^\alpha(t)}{N^3}+\dots\\
\beta^*(N,t)&\approx &p_0^\beta(t)+\frac{p_1^\beta(t)}{N}+\frac{p_2^\beta(t)}{N^2}+\frac{p_3^\beta(t)}{N^3}+\dots
\end{eqnarray}
For large $N$, the higher-order terms in the expansion are unimportant. As seen in Fig. 5 of the main text, there is good agreement between the cubic and quadratic fits, with the largest difference on the order 1\% for $\beta^*$. Furthermore, the cubic fits are in excellent agreement with TEBD results for large systems. Therefore, the parameters $p_0^{\alpha,\beta}(t)$ obtained from the cubic fit provide a smooth, accurate extrapolation of the optimal variational parameters to the thermodynamic limit, $N{\to}\infty$.


\section{Error Mitigation Techniques}

The Hamiltonian simulation via Trotterisation executed on Noisy Intermediate-Scale Quantum (NISQ) computers suffers from physical errors in addition to the algorithmic. Quantum error mitigation approaches are essential to exploit the quantum advantage of NISQ devices. Noise-aware quantum compilers, which translate the quantum algorithm into physical quantum circuits supported by the underlying quantum hardware, incorporate different error mitigation approaches into the compilation process. We apply several optimization layers to boast the fidelity of the quantum circuits, which can be integrated with the state-or-the-art quantum circuit compilation approaches (e.g. Quantum Information Software Kit (Qiskit~\cite{Qiskit})). We use Qiskit compiler to generate physical quantum circuits. This process involves optimizing the gate count and the circuit depth. Given the physically mapped and optimized quantum circuit, quantum gates are rescheduled based on the gate commutation rules~\cite{Optimiz_Q,farhi2014quantum} to reduce the quantum circuit error rates~\cite{saravanan2022pauli}. The objective is to push gates with very high error rates to later layers in the quantum circuit to limit their error impact on the circuit when the variation in the gate error rates is significant. The rescheduling algorithm maintains the circuit depth and therefore does not introduce new decoherene errors. This rescheduling algorithm can also create new optimization opportunities, which can further reduce the gate count. We also exploit the trade-off between the algorithmic errors and the physical errors of the trotterisation through the heuristic quantum search compiler~\cite{9259942}. This compiler constructs a quantum sub-circuit that realizes a unitary matrix approximately by searching for a sequence of quantum gates which form a unitary matrix such that the distance between the approximated and the target unitary matrices is within the accepted threshold. To ensure a reduction in the CNOT gate count, we iteratively increase the error threshold until we observe an improvement in the circuit gate count. The maximum threshold value that we consider is $10^-{2}$. The approximation can be applied to one or more trotter steps depending on the injected errors in each approximated subcircuit.
\bibliographystyle{apsrev4-1}
\bibliography{FQHE.bib}